\documentclass[12pt,preprint]{aastex}
\usepackage{cite}
\usepackage{graphicx}
\usepackage[header,title,page,titletoc]{appendix}
\usepackage{natbib}
\usepackage{hyperref}  

\shorttitle{22 $\mu$m Excess Candidates}
\shortauthors{Wu et al.}
\begin{document}
\title{Bright 22 $\mu$m Excess Candidates from WISE All-Sky Catalog and 
    Hipparcos Main Catalog}
\author{Chao-Jian Wu\altaffilmark{1,2,3}, Hong Wu\altaffilmark{1,3},
    Man-I Lam\altaffilmark{1,3}, Ming Yang\altaffilmark{1,3}, 
    Xiao-Qing Wen\altaffilmark{3}, Shuo Li\altaffilmark{4},
    Tong-Jie Zhang\altaffilmark{2}, \& Liang Gao\altaffilmark{1}}

\altaffiltext{1}{National Astronomical Observatories, 
                Chinese Academy of Sciences, Beijing 100012, P.R. China}
\altaffiltext{2}{Department of Astronomy, Beijing Normal University, Beijing,
                100875, P. R. China}
\altaffiltext{3}{Key Laboratory of Optical Astronomy, National Astronomical
                Observatories, Chinese Academy of Sciences, Beijing 100012,
                P.R. China}
\altaffiltext{4}{Department of Astronomy, Peking University, Beijing,
                100871, P. R. China}

\begin{abstract}
  In this paper we present a catalog which includes 141 bright candidates 
  ($\leq10.27$ mag, V band) showing the infrared (IR) excess at 22 $\mu$m. 
  Of which, 38 stars are known IR excess stars or disk, 23 stars are double or 
  multiple stars and 4 are Be stars. While the 
  remaining more than 70 stars are identified as the 22 $\mu$m excess 
  candidates in our work. The criterion of selecting candidates is 
  $K_s-[22]_{\mu m}$.
  All these candidates are selected from \emph{WISE} All-sky data 
  cross-correlated with \emph{Hipparcos} Main Catalog and the likelihood-ratio 
  technique is employed. Considering the effect of background, we introduce the 
  \emph{IRAS} 100 $\mu$m level to exclude the high background. We also 
  estimated the coincidence probability of these sources.
  In addition, we presented the optical to
  mid-infrared SEDs and optical images of all the candidates, 
  and gave the observed optical spectra of 6 stars with NAOC's 2.16-m telescope. 
  To measure for the dust amount around each star, 
  the fractional luminosity is also provided. We also test whether our method 
  of selecting IR excess stars can be used to search for extra-solar
  planets, we cross-matched our catalog with known IR-excess stars
  having planets but none is matched. Finally, we give the
  fraction of stars showing IR-excess for different spectral type of
  main-sequence stars.
\end{abstract}

\keywords{infrared: stars-planetary systems - stars: formation -
planetary systems: protoplanetary disks}

\section{INTRODUCTION \label{intro}}
The nature of IR excess is still uncertain: it may be produced by
protostars \citep{thompson1982} or the surrounding dust disk
\citep{gorlova2004, gorlova2006, rhee2007, hovhannisyan2009,
  koerner2010, wu2012}, or from giant stars, and it may also be due to
M dwarfs, brown dwarfs \citep{debes2011}. However, the IR excess could
also come from the companion star, background galaxy, background nebula
or interstellar medium or random foreground object, not from the object 
itself\citep{ribas2012}.

Since the first discovery of the debris disk around Vega by infrared excess,
\citep{aumann84}, IR excess is a useful tool to search debris disk.
To date, many works on searching for stars with IR excess emission
have been conducted. Most of the samples used in the previous work
were selected from \emph{IRAS}, \emph{Infrared Space Observatory
  (ISO)} and \emph{Spitzer Space Telescope} \citep{rhee2007,
  lagrange2000, zuckerman2001, decin2003}, and the searching
wavelength was focused at 60 $\mu$m or 70 $\mu$m. E.g.,
\citet{rhee2007} identified 146 stars that show the excess emission at
60 $\mu$m by cross-correlating \emph{IRAS} catalogs with
\emph{Hipparcos} stars, and 33 stars were found to have debris
disks. In addition, several other papers published between 2004 and
2005 reported that many Vega-like stars detected by \emph{Spitzer} at
70 $\mu$m have not been detected at 60 $\mu$m by \emph{IRAS} and
\emph{ISO} \citep{meyer2004, chen2005b, beichman2005, low2005,
  kim2005}.

Some works were also done at shorter wavelengths, e.g., 24
$\mu$m. Especially after the launch of the \emph{Spitzer} Space
Telescope \citep{werner2004}, many \emph{Spitzer} sub-programs were
carried out. For example, some wide-field surveys from the Multi-band
Imaging Photometer (MIPS) \citep{rieke2004} were performed at three
mid-to-far infrared bands (24, 70 and 160 $\mu$m). Many stars with 24
$\mu$m excess are studied using the MIPS's 24 $\mu$m
database. \citet{low2005} found 4 out of 24 stars in the 810 Myr old
TW Hya association shows 24 $\mu$m excess. \citet{young2004} found
several stars with 24 $\mu$m excess from the cluster NGC 2547.
\citet{gorlova2004, gorlova2006} found stars with 24 $\mu$m excess
from the open cluster M47 and Pleiades cluster using the selection
criterion $K_S-[24]_{vega}\geq0.44$. \citet{su2006} reported that the
24 $\mu$m excess occurrence rate is about 32\% by studying 160 A-type
main-sequence stars. 
At high galactic latitudes, \citet{wu2012} found eleven 24 $\mu$m excess stars with older age.

With the release of the Wide-field Infrared Survey Explorer (WISE;
\citeauthor{wright2010}~\citeyear{wright2010}) all sky data, the
observations at 22 $\mu$m will undoubtedly provide an opportunity to
search for more IR excess stars in the whole sky \citep{wu2012}. Some related 
work have been published. \citet{kennedy2012} described a search for IR excess 
stars from \emph{Kepler} and \emph{WISE} and concluded that the excesses in the 
\emph{Kepler} field are mainly due to high background level. \citet{lawler2012} 
studied the dust emission around more than 900 \emph{Kepler} exoplanet 
candidates using \emph{WISE} data and they found 8 candidates with IR excess. 
\citet{morales2012} studied the dust of 591 planetary systems from Exoplanet 
Encyclopaedia as of 2012 January 31, 350 can be detected by \emph{WISE} and 9 
stars have mid-IR excess. \citet{avenhaus2012} searched IR excess mainly for M 
stars. In our work, we focus on the bright stars and the observed information 
at 22$\mu$m. More information about WISE will be described in Section
\ref{wise}.

In order to study the properties of the IR excess stars in more
detail, the information of WISE is not enough, and we need more
observed quantities, e.g., optical data, distance, spectral type and
so on. So we choose the \emph{Hipparcos} Main Catalog for
cross-correlating with WISE due to its highest photometric precision
and distance information. The Hipparcos Catalog, one of the two
major stellar catalogs resulting from the ESA’s Hipparcos space
astrometry project, was completed in August 1996, and published in
June 1997 (ESA 1997). In Section \ref{hipparcos}, detailed information
of \emph{Hipparcos} are presented and the reason of only using the
\emph{Hipparcos} Main Catalog is also explained.

In the previous study, similar works were purely based on the
\emph{WISE} and SDSS DR7 \citep{debes2011}or \emph{IRAS} and
\emph{Hipparcos} catalogs \citep{rhee2007}. In this work, we firstly
use the all-sky WISE data to search for bright IR excess stars by matching
with \emph{Hipparcos} Catalog.
Generally, color-color diagram is a useful tool for
detecting IR excess \citep{hoard2007, wellhouse2005, wachter2003}.

In this paper, we describe the \emph{WISE} all sky data,
\emph{Hipparcos} Catalog, candidate selection criterion, source
identification method and optical observations in Section
\ref{candXobs}. In Section \ref{results}, we classified the IR
excess stars, analyzed their infrared properties of and presented their 
spectral energy distributions (SEDs) and the optical images. The conclusion and 
summary are presented in Section \ref{summary}.

\section{CANDIDATES SELECTION AND OBSERVATIONS \label{candXobs}}
\subsection{WISE All-Sky Catalog \label{wise}}
The WISE satellite was launched on December 14, 2009. It mapped the
sky at $3.4$, $4.6$, $12$, and 22 $\mu$m (W1, W2, W3, W4) with an
angular resolution of $6''.1$, $6''.4$, $6''.5$, and $12''.0$ in the
four bands, respectively, achieving 5$\sigma$ point source sensitivities
better than 0.08, 0.11, 1 and 6 mJy in the four bands in unconfused regions
on the ecliptic plane\citep{wright2010}. The all-sky data were released on
March 14th 2012 and it includes all the data taken during the WISE full
cryogenic mission phase, from January 7th 2010 to August 6th 2010, which were
processed with improved calibrations and reduction
algorithms. Released data products include an Atlas of 18,240 image
sets, a source catalog containing positional and photometric
information for over 563 million objects detected on the WISE images.
It supersedes the Preliminary Data which was released in April,
2011\footnotemark[1].
\footnotetext[1]{http://wise2.ipac.caltech.edu/docs/release/allsky/}
The mission of \emph{WISE} has several main goals, such as taking a
census of cool stars and brown dwarfs close to the Sun, probing the
dustiest galaxies in the universe, and cataloging the Near Earth
Object population \citep{wright2010, debes2011}. It will also provide
crucial information on the IR sky at a sensitivity 100 times better
than that of \emph{IRAS}.However, the WISE team found an overestimate in brightness 
in the 4.6 $\mu$m (W2) band \footnotemark[2] and the bias can be reach nearly 1
mag \citep{tisserand2012}.
\footnotetext[2]{http://wise2.ipac.caltech.edu/docs/release/allsky/expsup/sec6\_3c.html}
In our work, we also found the existence of the bias on bright sources
at 4.6 $\mu$m band. It will be described in Section \ref{ccdiag}.

In this paper, we firstly selected the all-sky sources from the WISE
All-Sky data catalog with the criterion $\mathrm{S/N\geq20}$ at W4 (22
$\mu$m) band, they contain positional and photometric information
including J, H, $\rm{K_s}$ bands of 2MASS and $3.4$, $4.6$, $12$, and 22 $\mu$m bands
(W1, W2, W3, W4) of WISE. After filtering with this criterion we
obtained a catalog with 971,148 sources. In the next step, it will
be used to cross-correlated with the \emph{Hipparcos} Main Catalog.

\subsection{Hipparcos Main Catalog \label{hipparcos}}
ESA's Hipparcos space astrometry mission was a pioneering European
project. It was launched in August 1989 and has successfully observed
the celestial sphere for 3.5 years before the operation was ceased in March
1993, and its scientific goal was to provide positions, proper
motions, and direct distance of stars near the solar system in order
to study the physical properties, stellar structure and evolution of
stars \citep{perryman1997, perryman1995}. The \emph{Hipparcos} Main
catalog was generated from these observations by the main instrument
and it includes 118,218 stars charted with the highest
precision. Also, an auxiliary star mapper scanned many more
stars with lesser accuracy, which was included in the Tycho Catalog
with 1,058,332 stars. The Tycho 2 Catalog, completed in 2000, brings
the total to 2,539,913 stars, and most of sources are bright stars
with an apparent magnitude of 11. It provides the information on the
position, proper motion and direct distance estimate for over 100,000
stars in the solar neighborhood \citep{perryman1997}. So many observed
parameters of stars in Hipparcos Catalog provide us enough
information to study their physical characteristics. In this work, we
used only the \emph{Hipparcos} Main Catalog because of its highest
precision. Meanwhile, it provided trigonometric parallaxes and proper
motions for more than 100,000 stars with errors 1$\sim$2 mas.

\subsection{Cross-Correlation \label{cross}}
As described in the sections above, we cross-correlated the selected
WISE sources with the \emph{Hipparcos} Main Catalog. The matching
radius we used here is $6''$, which is consistent with the
Full-Width-at-Half-Maximum (FWHM) of the WISE's PSF at 3.4 $\mu$m
\citep{wright2010}.  The cross-matched catalog was obtained, and it
contains about 66,667 sources. But these sources can't be used directly. 
They should be filtered with full WISE and 2MASS photometric information. 
Moreover, we also select the \emph{Hipparcos} sources with threshold 
$\delta_{plx}<0.1$ meaning distance accuracy should be better than 
$10\%$ and with a photometric error $\delta_{B-V}<0.025$ \citep{perryman1995}. 
Moreover, we should delete those saturation sources at $K_s$, $W3$ and $W4$ 
band. In fact, the photometric error at $K_s$ band should also be noted, 
because $K_s$ will be used as the criterion of selecting IR excess. To  
ensure the statistical accuracy (See Section \ref{selection}), only those with 
$\sigma\leq0.1$ at $K_s$ band are selected. That is to say, 7,624 sources 
are used to search IR excess stars in our work.

\subsection{Candidates Selection \label{selection}}
\subsubsection{$K_s-[22]$ Criterion \label{Ks_22}}
In this section, we will describe how to identify the IR excess candidates 
from the 7,624 samples (black dots in Figure \ref{allsky_sfr} ). They all 
contains
multi-band information, e.g., parallax value, spectral type and so
on. Once we find the IR excess stars from the cross-correlated
catalog, we can make a detailed study to the selected sources with
so much observational information.

\begin{figure}[h!tb]
\begin{center}
\resizebox{15cm}{!}{\includegraphics{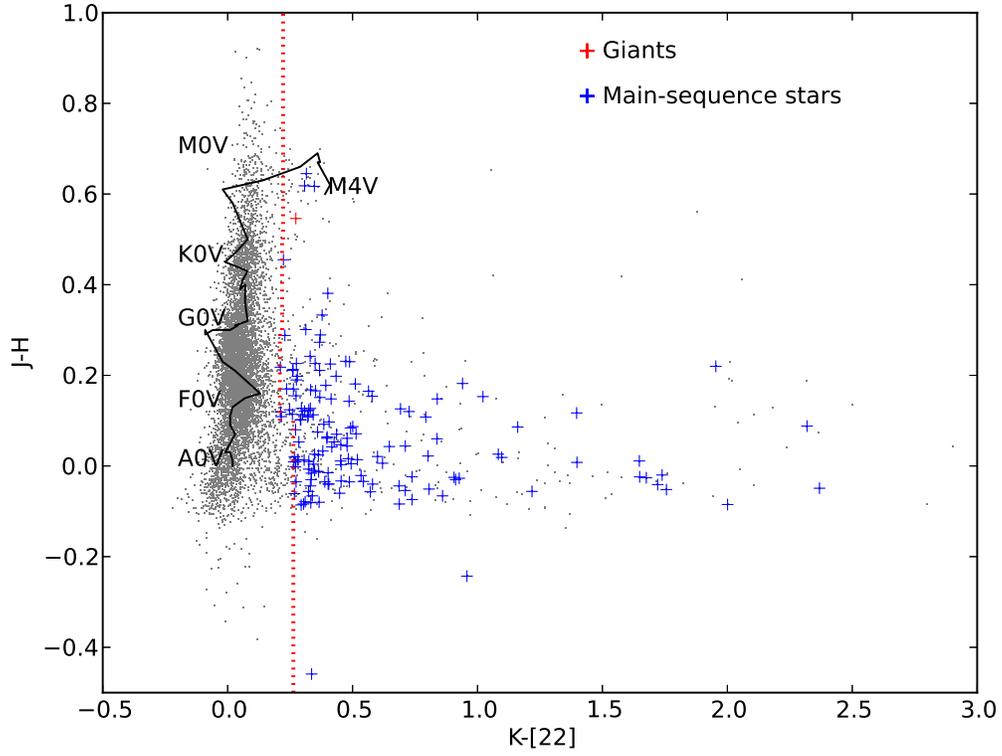}}
\end{center}
\caption{Diagram of $J-H$ vs $K_s-[22]$. The distribution of main-sequence 
    stars and Giants are plotted as blue and red plus symbol respectively.
    The black solid line shows the normal dwarf stars labeled with the
    corresponding spectral types. The red dotted line gives our criterion for
    selecting the 22 $\mu$m excess sources. 
\label{k_22}}
\end{figure}

\begin{figure}[h!tb]
\begin{center}
\resizebox{15cm}{!}{\includegraphics{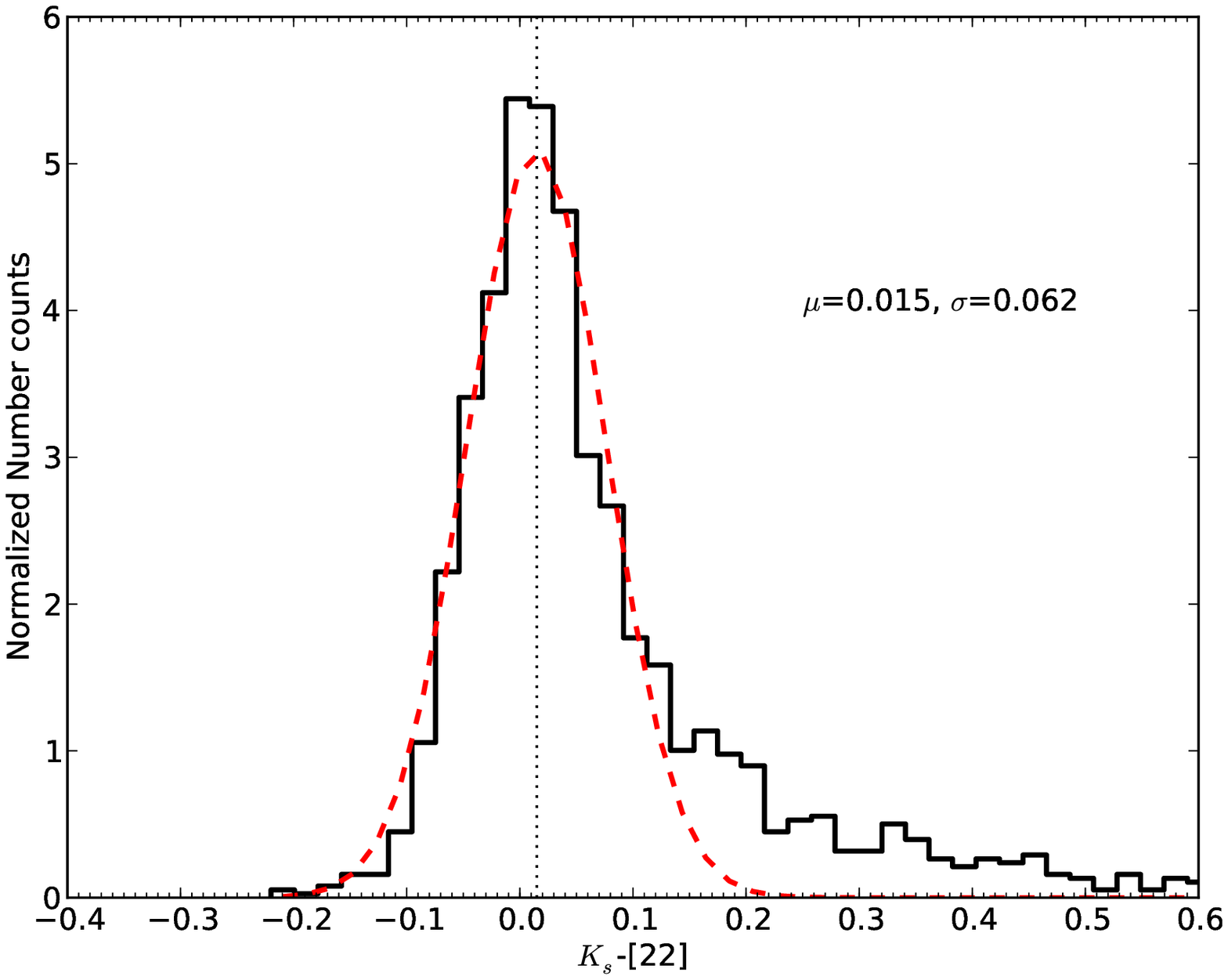}}
\end{center}
\caption{Goodness of fit for sources with $J-H\le0.1$. The criterion is 
$K_s-[22]_{\mu m}\geq0.26$.\label{hist:jhlt0.1}}
\end{figure}

\begin{figure}[h!tb]
\begin{center}
\resizebox{15cm}{!}{\includegraphics{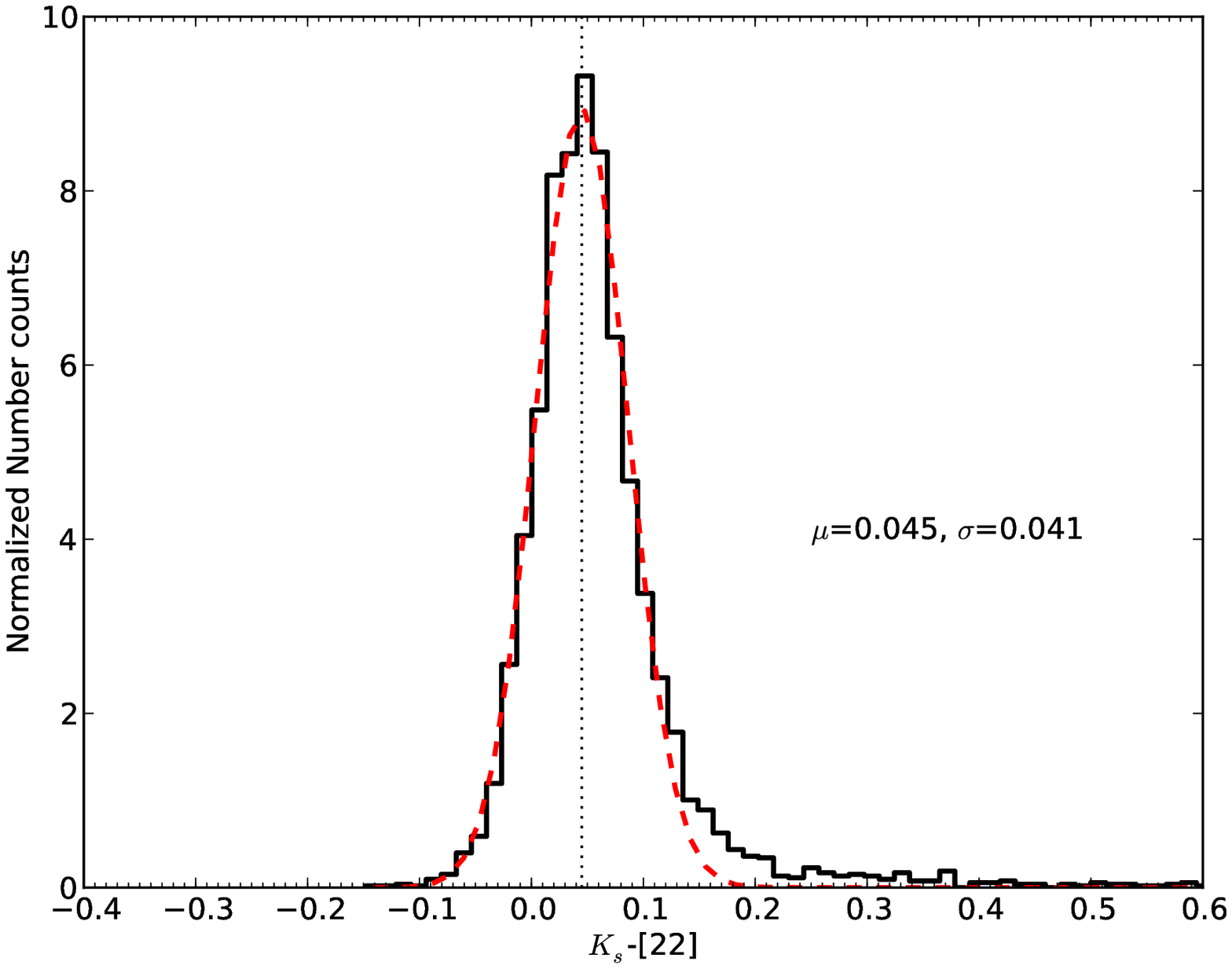}}
\end{center}
\caption{Goodness of fit for sources with $0.1<J-H\le0.3$. The criterion is 
$K_s-[22]_{\mu m}\geq0.21$.\label{hist:jh0.10.3}}
\end{figure}

\begin{figure}[h!tb]
\begin{center}
\resizebox{15cm}{!}{\includegraphics{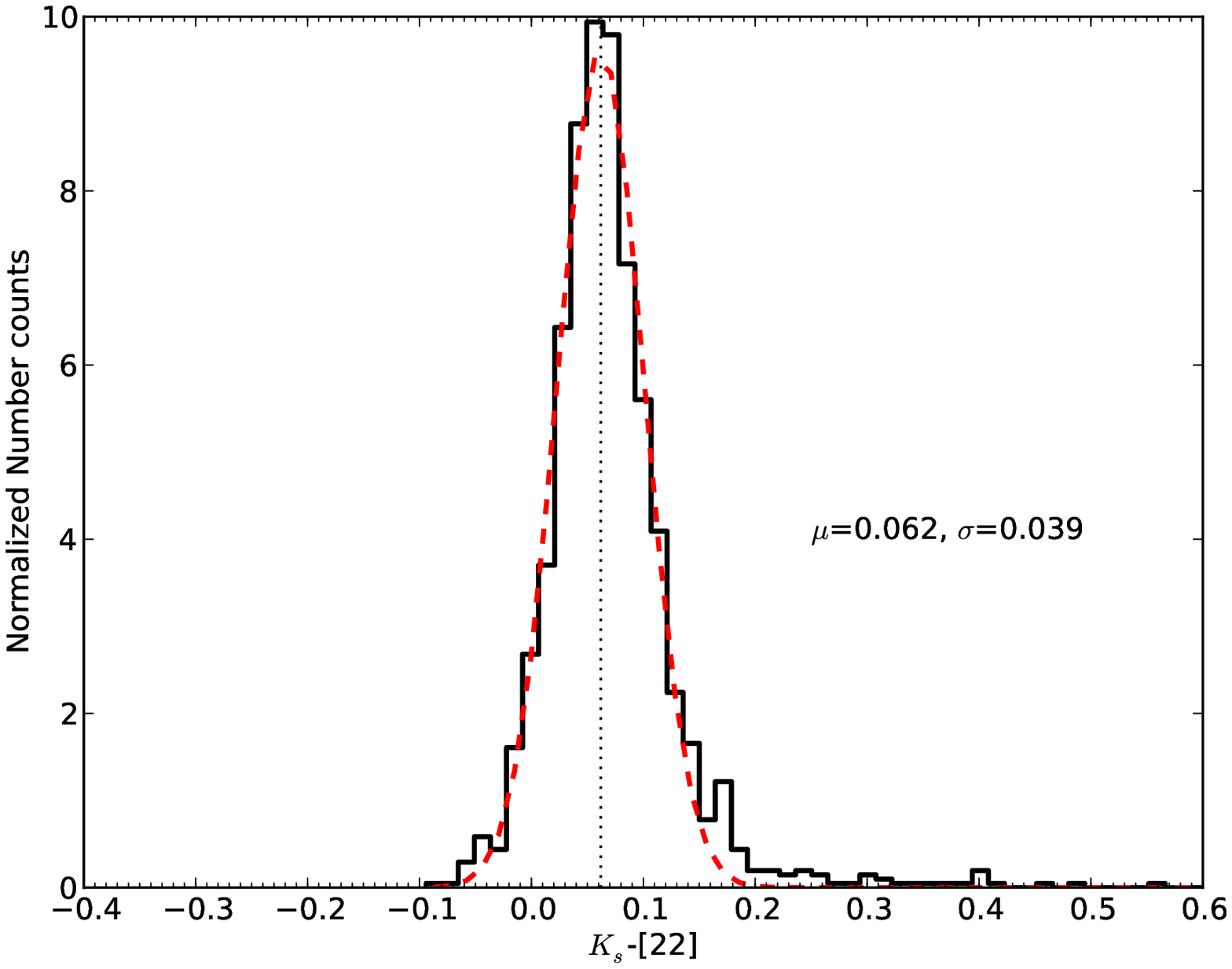}}
\end{center}
\caption{Goodness of fit for sources with $0.3<J-H\le0.5$. The criterion is 
$K_s-[22]_{\mu m}\geq0.22$.\label{hist:jh0.30.5}}
\end{figure}

\begin{figure}[h!tb]
\begin{center}
\resizebox{15cm}{!}{\includegraphics{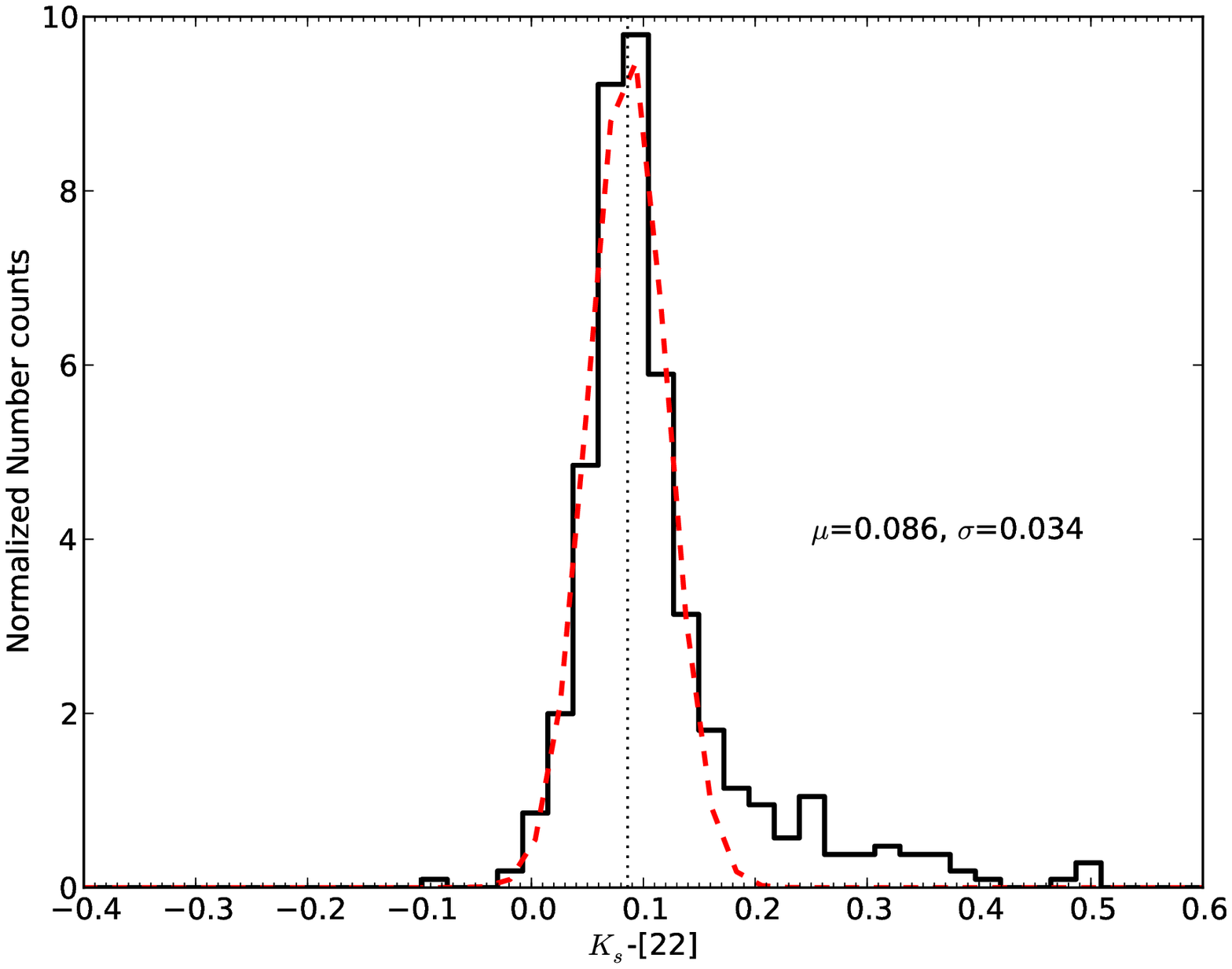}}
\end{center}
\caption{Goodness of fit for sources with $J-H>0.5$. The criterion is            
$K_s-[22]_{\mu m}\geq0.22$ \label{hist:jhge0.5}}.
\end{figure}

\citet{gorlova2004, gorlova2006} provided an approach on searching for 
IR excess in the mid-infrared band, and they used the criterion that
the mean $K_s-[24]_{vega}$ (here $[24]$ means the vega magnitude at 24 $\mu$m, 
$[22]$ has the same meaning) value should be greater than 0.33 at a 3$\sigma$ 
confidence level ($0.33=3\times 0.11$, where 0.11 is the 1$\sigma$ value). 
While in \citet{hovhannisyan2009}, the criterion was changed slightly to be 
$K_s-[22]\geq0.2$, but in this paper, there is an assumption that all the 
stars have $K_s-[24]=0$. However, this assumption is invalid for our sources. 
WISE team shows that there is calibration offset relative to 2MASS K-band, 
in other word, the WISE-K color is not zero. So we have to redefined the 
criterion. Similar to \citet{gorlova2004}, the histogram of $K_s-[22]$ can 
can help us to define the criterion. From Figure \ref{k_22}, we can see 
that the points along y axis show different scatter. So we divided 
our samples into four parts to do statistics following $J-H\le0.1$, 
$0.1<J-H\le0.3$, $0.3<J-H\le0.5$ and $J-H>0.5$, respectively. The results 
are shown in Figure \ref{hist:jhlt0.1}, Figure \ref{hist:jh0.10.3}, 
Figure \ref{hist:jh0.30.5}, and Figure \ref{hist:jhge0.5}. The 
histogram of $K_s-[22]$ colors can be 
described by a Gaussian centered at $K_s-[22]=0.015$ mag with $\sigma=0.062$ 
mag for $J-H\le0.1$, $K_s-[22]=0.045$ mag with $\sigma=0.041$ mag for 
$0.1<J-H\le0.3$, $K_s-[22]=0.062$ mag with $\sigma=0.039$ mag for 
$0.3<J-H\le0.5$ and $K_s-[22]=0.086$ with $\sigma=0.034$ mag for $J-H>0.5$ 
sources, respectively. We therefore define IR excess stars as those lying 
redward of $K_s-[22]=0.015+4\sigma=0.26$ for $J-H\le0.1$, 
$K_s-[22]=0.045+4\sigma=0.21$ for $0.1<J-H\le0.3$, 
$K_s-[22]=0.062+4\sigma=0.22$ for $0.3<J-H\le0.5$ and 
$K_s-[22]=0.086+4\sigma=0.22$ for $J-H>0.5$. As shown in Figure \ref{k_22}, 
495 sources (those locate at the right side of red dotted line) have 22 
$\mu$m excess by using our criterion. 

However, the 495 sources can not be identified as really IR excess stars. The 
excess may be from nearby bright stars or background. So we introduce another 
two criterion to exclude the fake candidates.

\subsubsection{Likelihood Ratio \label{lr}}
We used the likelihood-ratio (LR) technique to identify the WISE IR
excess sources. Its principle is to accept the nearest optical
source. The LR method was first used by \citet{richter1975} and
defined as in \citet{sutherland1992}:
\begin{equation}
    L=\frac{q(m)f(r)}{n(m)}
\end{equation}
where $f(r)$ is the radial probability distribution function of the positional
errors with separation in arcseconds $r$, given by \citet{smith2011}
\begin{equation}
    f(r)=\frac{1}{2\pi\sigma_{pos}^2}\mathrm{exp}(-r^{2}/{2\pi\sigma_{pos}^2})
\end{equation}
in which $\sigma_{pos}$ is the uncertainty for the position, while $n(m)$ and
$q(m)$ correspond to the surface density per magnitude and the probability
distribution function, respectively.

In this work, for a WISE candidate with a magnitude of $m$ ($V$ mag from
$Hipparcos$) at an angular separation $r$ from a given optical source,
the LR is deﬁned as the ratio of the probability of the WISE object
being the true counterpart of the optical source
\citep{ciliegi2003}. We assume that the probability distribution of
angular separations follows a Gaussian distribution, as given by
\citet{sutherland1992}, we can rewrite the LR as:
\begin{equation}
    L=\frac{Q(\leq m_i)exp(-r^{2}/2)}{2\pi\sigma_{pos}^2 n(\leq m_i)}
\end{equation}
$Q(m)$ is the expected magnitude distribution of counterparts. It is given by
\begin{equation}
    Q=\int_{-\infty}^{m_{lim}}q(m)dm
\end{equation}

Generally, the positional uncertainty should depend on the
signal-to-noise ratio ($S/N$) and on the FWHM. So we use the results
derived by \citet{ivison2007}, which gave
\begin{equation}
    \sigma_{pos}=0.6\frac{\mathrm{FWHM}}{\mathrm{S/N}}
\end{equation}

\begin{figure}[h!tb]
\begin{center}
\resizebox{15cm}{!}{\includegraphics{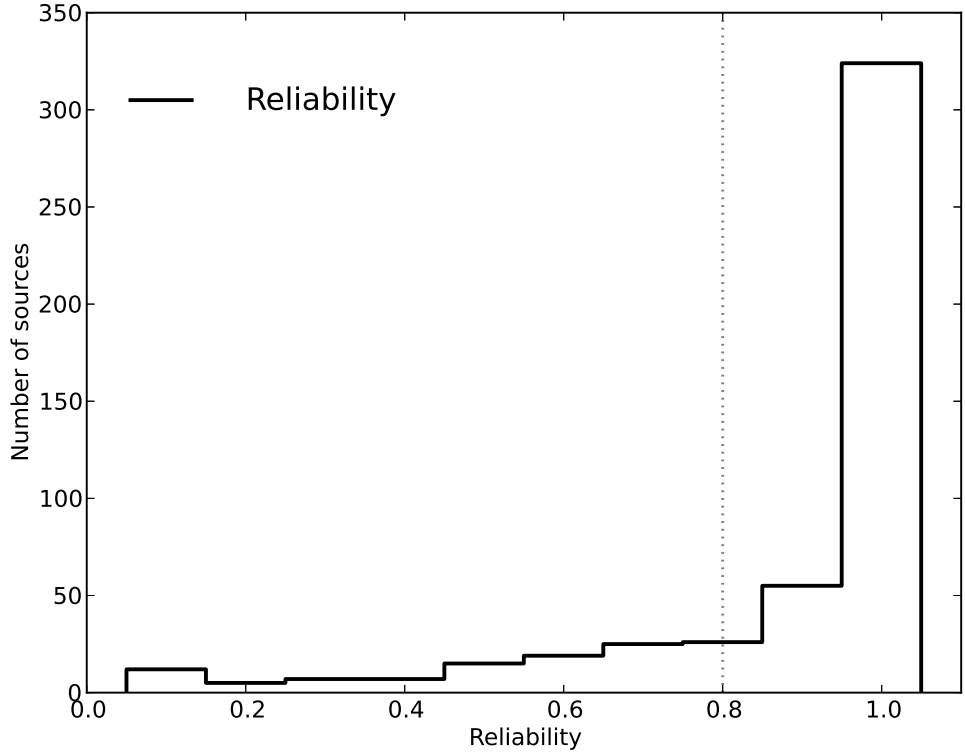}}
\end{center}
\caption{Reliability histogram of all the candidates. The dotted line shows
    the selection threshold. Those with reliability $R<0.8$ are excluded.
    \label{reliability}}
\end{figure}

Given LR, so we can define the reliability $R_i$ for the
$i$th counterpart again following \citet{sutherland1992}:
\begin{equation}\label{relia}
    R_i=\frac{L_i}{\Sigma_i L_i+(1-Q)}
\end{equation}
where $Q$ stands for the probability that the counterpart of the
source is above the limiting magnitude. In \citet{mainieri2008}, they
pointed out that values of $Q$ in the range 0.5-1.0 will make no
significant difference in the results. So we also choose $Q=0.8$ in
this work.  We used equation \ref{relia} to calculate the reliability
of all candidates and then we select those sources with reliability
$R\geq0.8$ and leave 378 sources in the IR excess catalog from 
Section \ref{selection}. The reliability distribution of these
sources is shown in Figure \ref{reliability}.

\subsubsection{The Contamination of Background}
\begin{figure}
\begin{center}
\resizebox{15cm}{!}{\includegraphics{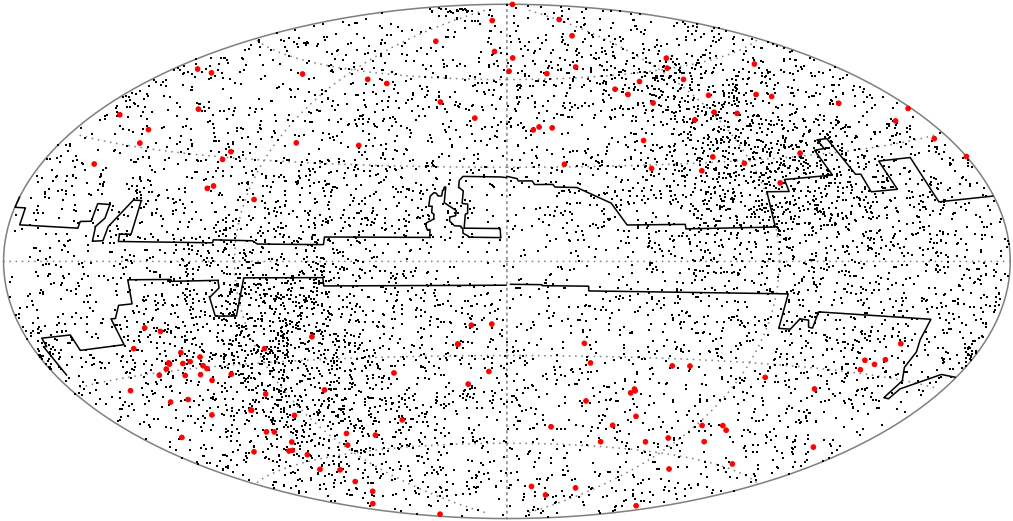}}
\end{center}
\caption{Distribution of IR excess stars. We show the distribution of matched
    catalog from \emph{WISE} All-Sky Catalog and \emph{Hipparcos} Main
    catalog in this
    figure (Gray dots); and the red points are the 141 IR excess candidates 
    selected in this work. The region between black solid line is the star
    formation region. They all show in a galactic Aitoff projection.
    \label{allsky_sfr}}
\end{figure}

\begin{figure}[h!tb]
\begin{center}
\resizebox{15cm}{!}{\includegraphics{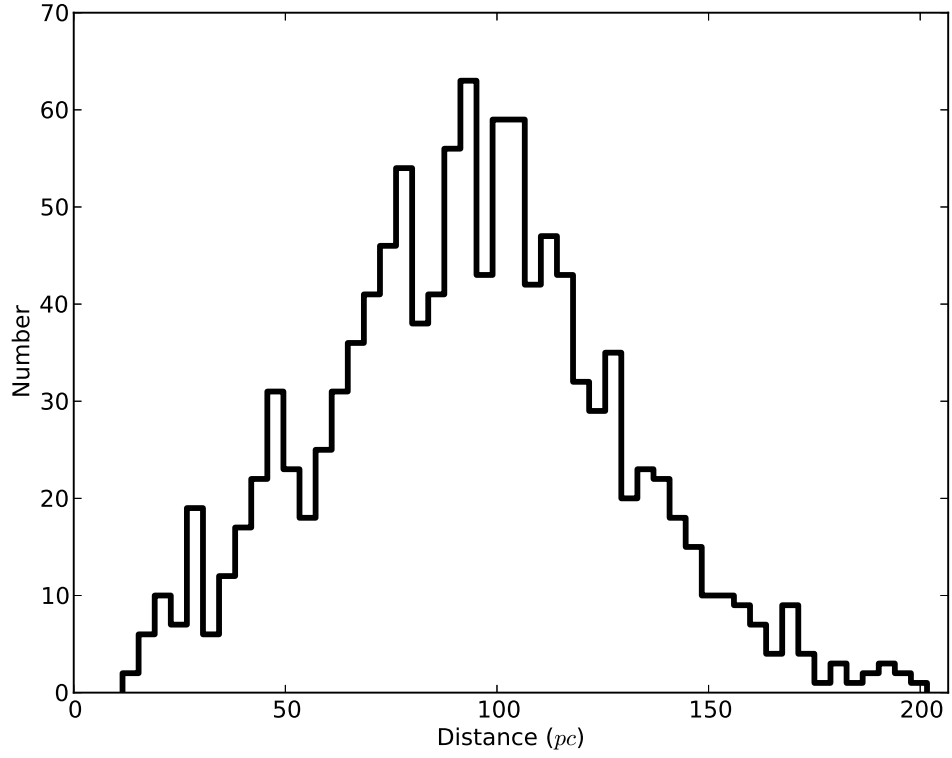}}
\end{center}
\caption{Distance distributions of all the candidates. The distance ranges
    within 200 $\mathrm{pc}$. The distance is so close that most of them locate
    in front of star formation region. \label{distance}}
\end{figure}

With the parallax value provided by \emph{Hipparcos}, the distance 
distribution is shown in Figure \ref{distance}. The distance of a star 
is a very important parameter. Not only it can be used for luminosity 
classification, but it can be used to determine whether the star is 
located in the star formation region. From Figure \ref{distance}, we 
can tell that almost all of these candidates have distances within 200pc. 
The distance is so close that most of them are located in the front of star 
formation region. By comparing with the distance of several nearby star 
formation regions and molecular clouds \citep{bertout1999}, like Taurus 
\citep{kenyon1994} and Ophiuchus \citep{knude1998, mamajek2008}, we found 
that none of our candidates were located in these nearby star formation 
regions or molecular clouds. They may just affect these sources as background. 
\citet{kennedy2012} shown that the $IRAS$ 100$\mu$m background level should be 
lower than 5$\mathrm{MJysr^{-1}}$. Of the 378 candidates left in Sec \ref{lr}, 
141 remain after this cut. One thing should be noted, all the 141 sources 
(red dots in Figure \ref{allsky_sfr}) 
locate in high latitude ($|l|>10^o$). That is to say, the star formation 
regions affect slightly.

Figure \ref{allsky_sfr} shows the Aitoff projection in the galactic
coordinate. The molecular cloud and interstellar medium are located in
the region between two black solid lines
\citep{dame2001}. For simplicity, those located in the high latitude
are not plotted, which don't have big impact on our results.  From
Figure \ref{allsky_sfr}, we found that hardly any of these stars locates 
in the galactic disk because of high IR background. 

\subsection{Optical Observations \label{observations}}
Using the selected sources from the observations of \emph{Hipparcos}
as described in section \ref{selection}, we
can obtain the Hertzsprung-Russell Diagram (HRD). From these candidates, we 
choose 6 stars ((red asterisk in Figure \ref{mainXgiant})) using for spectral 
observation and they all located in the main sequence of HRD. The detailed 
informations of observation are described bellow.

The optical spectra were obtained with NAOC's 2.16-m telescope at Xinglong, 
Beijing in Jan. 2012. The attached spectrograph is obtained with the grism G7 
of BFOSC (Beijing Faint Object Spectrograph and Camera), and the spectrograph 
covers the wavelength range from $\mathrm{3870 \AA}\sim\mathrm{6760 \AA}$ . 
The exposure times were short because they are all bright stars. The detail 
informations of the spectral observations are listed in Table \ref{obs_stars}.

\begin{figure}[h!tb]
\begin{center}
\resizebox{15cm}{!}{\includegraphics{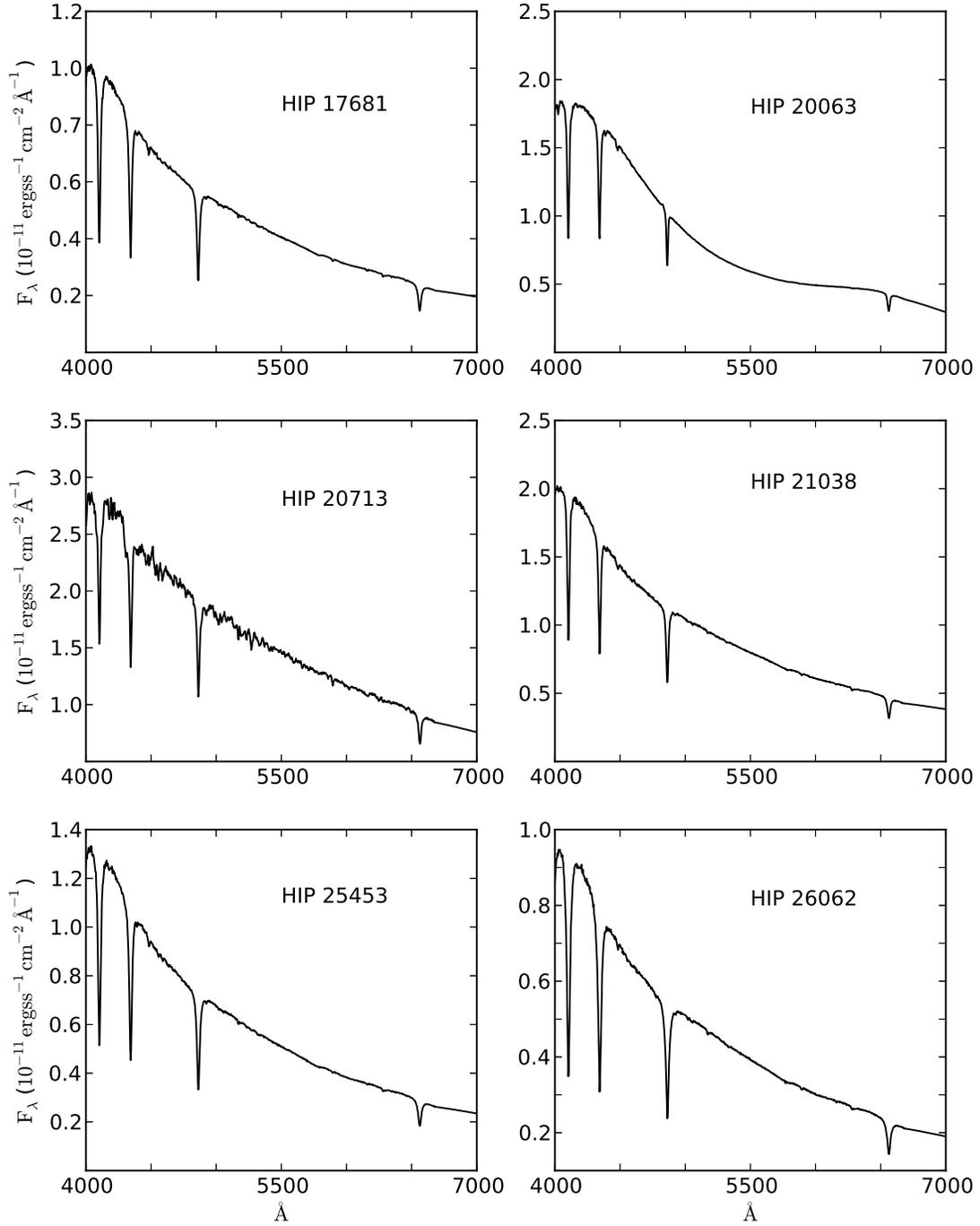}}
\end{center}
\caption{The observed optical spectra of six 22 $\mu$m excess stars. All
    stars have high $\mathrm{S/N}$ and present the main-sequence star features,
    covering the spectral types from B8 to F0.
    \label{spec}}
\end{figure}

All these spectral data were reduced by the standard procedures with IRAF
packages, which include overscan correction for BFOSC only, bias subtraction,
flat-field correction. The Fe/Ar lamps were used for the wavelength calibration
of BFOSC spectra. The standard stars used for the flux calibration at each night
are Feige25, HZ14, EG247, GD71, respectively. All resulting spectra are shown in
Figure \ref{spec}. The spectral classifications are also given in Table
\ref{obs_stars}, which is consistent with the spectral type given by
\emph{Hipparcos} Main catalog. The six stars are all the main-sequence
dwarf stars with the spectral types from B8 to F0.

\section{RESULTS AND DISCUSSION \label{results}}

\subsection{Notes on Catalog \label{notes}}
Then we present a catalog for our selected IR excess candidates, which 
containing the information provided in the \emph{WISE} and \emph{Hipparcos} 
Main Catalog. This catalog contains 38 known IR excess stars or debris disk 
candidates, 23 double or multiple stars, 12 variable stars and 4 Be stars. 
While the remaining more than 70 stars are identified as the 22 $\mu$m excess 
candidates in our work. 

\subsubsection{Catalog of IR Excess Stars \label{catalog}}
The parameters of the catalog are list bellow: The \emph{Hipparcos}
name, \emph{Hipparcos} RA and DEC (in the units of degree, J2000),
spectral type (given by \emph{Hipparcos}), the luminosity ratio
(calculated in Section \ref{fd}), photometric magnitude at optical
$B$, $V$ and $I$ bands, $J$, $H$, $K_s$ of 2MASS and four band of
\emph{WISE} and $K_s-[22]$ (mag, used as criterion for searching IR
excess). The photometric magnitude uncertainty of each band is also
listed. The Vega magnitude system has been used. 
The stars, which could be contaminated by the nearby stars that can not be 
excluded using likelihood-ratio method, are checked by optical images and 
marked in the last column.

The whole catalog has 141 IR excess candidates and each contains 18 columns.
A summary of the column information is given in Table
\ref{tab_sum}. They are 
presented in Table \ref{tab_main} for different types of stars. 

%
\subsubsection{Classification of IR excess stars \label{classification}}
\begin{figure}[h!tb]
\begin{center}
\resizebox{15cm}{!}{\includegraphics{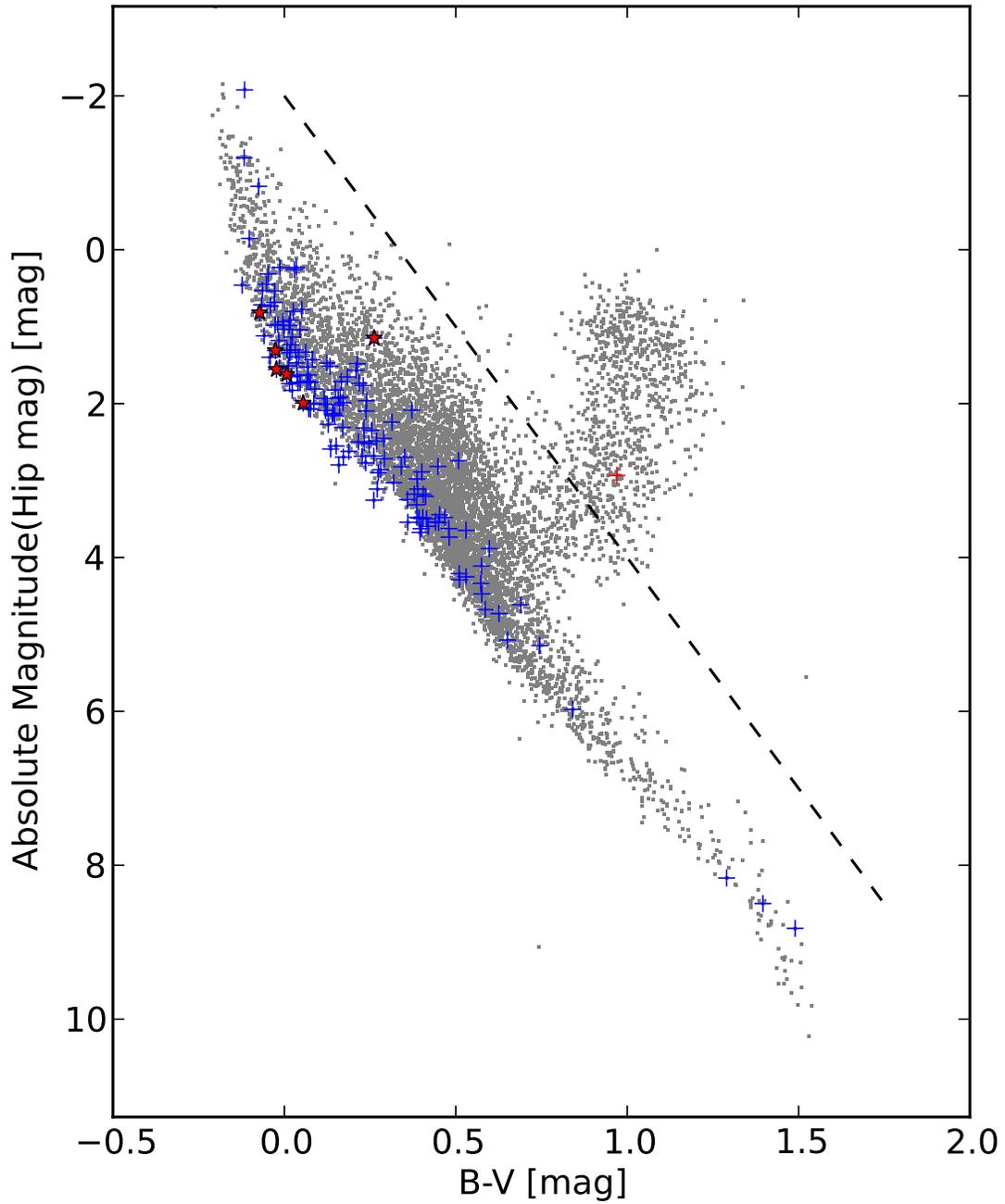}}
\end{center}
\caption{HRD of matched sources, main-sequence stars and giant
  stars. As shown in this figure, gray is the matched Hippocas main
  catalog with WISE, blue represents the main-sequence stars with IR
  excess and red represents the giants. The 6 red asterisks are observed by 
  2.16m Telescope. Main-sequence stars and giants are
  separated by the criterion $M_v>6.0(B-V)-2.0$ (indicated with dashed
  line). \label{mainXgiant}}
\end{figure}

\begin{figure}[h!tb]
\begin{center}
\resizebox{15cm}{!}{\includegraphics{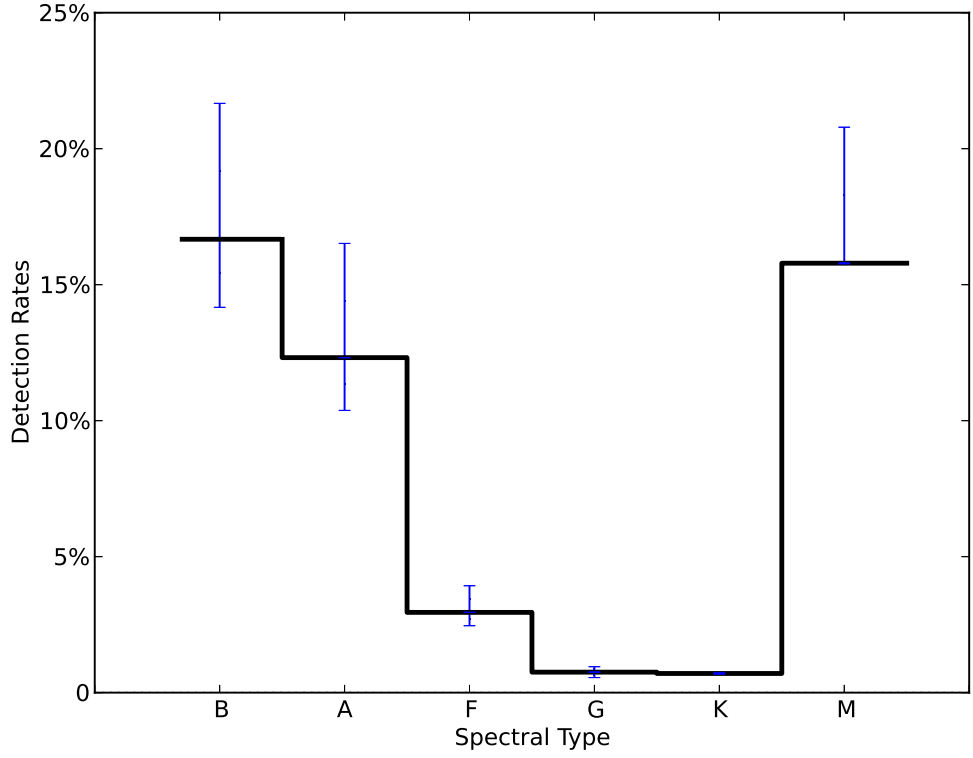}}
\end{center}
\caption{Detection rates of different spectral type stars showing IR excess, 
    the corresponding error bars (1$\sigma$) are also shown.
\label{spt_hist}}
\end{figure}

In the filtered catalog described in Section \ref{catalog}, 141 sources are 
included. All these stars show the IR excess feature at 22 $\mu$m. To study the 
principle of showing IR excess, we checked all these sources and classified 
them into the following two types.
\begin{itemize}
    \item Main-Sequence stars \\
        As shown in Figure \ref{mainXgiant}, some of them located in the giant
        stars region and most located in the main-sequence region. So we can
        separate the giant stars from the main-sequence stars in our
        catalog. The constraint $M_v>6.0(B-V)-2.0$ (dashed line in Figure
        \ref{mainXgiant}) are used to separate main-sequence stars (blue plus 
        symbol in Figure \ref{mainXgiant}) from this sample \citep{rhee2007}. 
        The separation
        between main-sequence stars and giants can help us to understand the
        possible different mechanism of the IR excess stars.
        There are 140 main-sequence stars which are presented in Table 
        \ref{tab_main}. All these main-sequence stars cover spectral type range
        from B5 to K0 (Figure \ref{spt_hist}), and most have luminosity type 
        \textrm{IV} and \textrm{V}. From the HRD (Figure \ref{spt_hist}), we 
        can see that almost all our candidates belong to the main-sequence 
        stars except one, so in order to avoid confusion, We remove the 
        luminosity type from \emph{Hipparcos} Main Catalog.
      \item Giants \\
        As described above, there are only one giant (red plus symbol in Figure 
        \ref{mainXgiant})which are listed in 
        Table \ref{tab_main}. Its spectral type is G5. 
\end{itemize}

\subsubsection{Contaminated Stars \label{contam}}
In Section \ref{classification}, we have classified them as
main-sequence stars and giants. 
Though most of our candidates don't locate in the galactic disc and have 
no background effect, the IR excess may still be contributed from the 
companion star. 
So they are all noted in Table \ref{tab_main}. 
We do this only by eyes. We plot the whole sample, as shown in Appendix 
\ref{appendixc}. From the optical and \emph{WISE} 22 $\mu$m images, almost 
all these contaminated stars have companion objects around. 
Because we can not confirm that whether these stars are affected by the 
surroundings, so they are marked and can be used for further study.

There are 11 contaminated stars found in our candidates. The 
contaminated stars can not be excluded perfectly by using the likelihood 
ratio method, because many surroundings of the contaminated stars are 
fainter than the center sources. That is why we must check them from 
optical and mid-IR images.

\subsection{The Coincidence Probability}
\begin{figure}[h!tb]
\begin{center}
\resizebox{15cm}{!}{\includegraphics{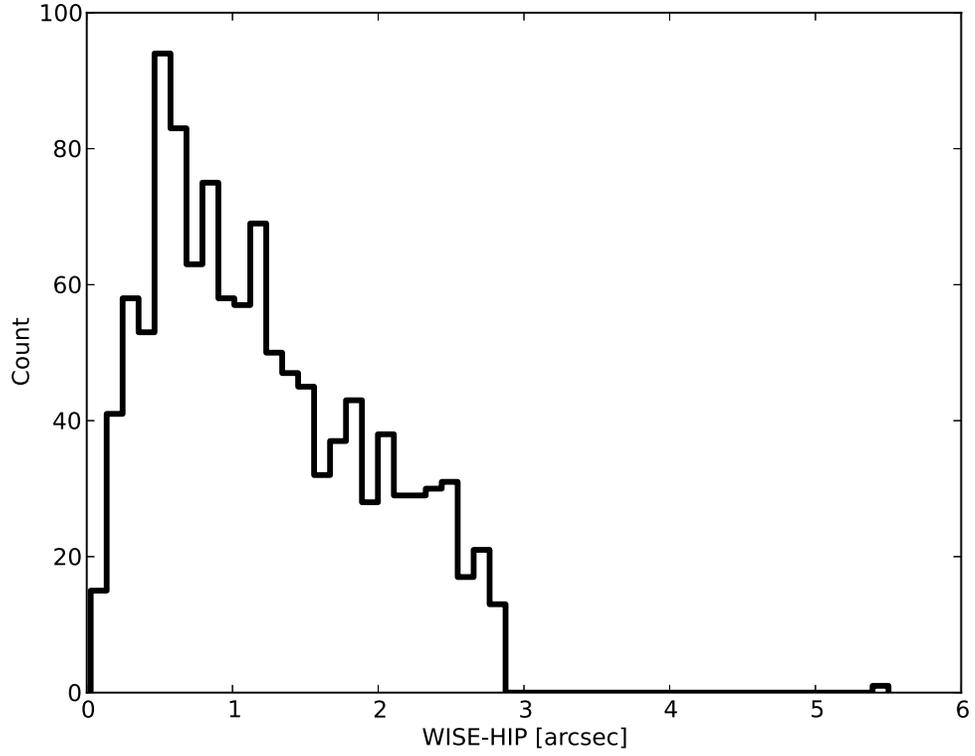}}
\end{center}
\caption{Center offset histogram of all the candidates between optical band and
    IR band. As shown in this figure, most
  stars except one have an uncertainty $\leq3$ arcsec. Thus we use 3 arcsec
  as the radius for calculating the coincidence probability. \label{radius}}
\end{figure}

Though we excluded the sources with the fuzzy features by using sources
identification method (Section \ref{lr}), it is still possible that some
background and distant galaxies coincide with the position of our the candidates
and contaminate the most radiation at 22 $\mu$m. Therefore, we need to estimate
the coincidence probability for each source. According to \citet{stauffer2005},
we first obtained the cumulative sources counts for those with 22 $\mu$m
magnitude brighter than the star itself. In fact, there is no need to calculate
all candidates' coincidence probabilities. We just want to give the probability
range. So we can roughly estimate the coincidence probability like this:
Firstly, we assume the \emph{WISE} all-sky data is evenly distributed. Then we
select the faintest stars (6.775 mag at 22 $\mu$m band with 
$\mathrm{S/N\geq20}$)
from the 141 candidates and estimate the cumulative sources counts with 22
$\mu$m magnitude less than $6.775$ in \emph{WISE} All-sky data catalog. The 
total number is about $2.5\times10^6$, which means about $7.6\times10^4$ per 
steradian. It corresponds to about one
source per $5.6\times10^5$ $\mathrm{arcsec}^2$. Considering the previous
matched radius of 6 arcsec. So this coincidence probability of background 22
$\mu$m source with magnitude less than 6.775 being close to the line of sight to
the faintest candidates is about 0.0002. Moreover, from Figure \ref{radius} we
can
see that almost all the center coordinate errors are $\leq3$ arcsec.
So we select 3 arcsec as the radius to calculate the coincidence probability.
Then each coincidence probability of the 141 candidates is 
$\leq5.0\times10^{-5}$. That is to say, within 3 arcsec radius, the
probabilities of meeting another source is only $0.00005$. In Section
\ref{hipparcos}, there are about 100,000 \emph{Hipparcos} sources used for
matching with \emph{WISE} catalog. We assume that the stars are evenly
distributed, so the coincidence sources are $\sim 5$. In fact, some of the
coincidence sources can be excluded with likelihood ratio method (Section
\ref{lr}).
The coincidence probability is so small that the star is almost impossible to
be contaminated which locates in the low-density star region (like the high
galactic latitude region). However, for those locate in the high density region,
we still can not exclude the possible contamination by interstellar medium or
background AGN \citep{stauffer2005}. So we should check the contaminated sources
from the optical images (Section \ref{contam}).

\subsection{Color-Color Diagram \label{ccdiag}}
Figure \ref{k_22} shows the $J-H$ versus $K_s-[22]$ diagram for all matched stars
from \emph{WISE} and \emph{Hipparcos} Main catalog. The gray dots are those 
have no excess at 22$\mu$m band and the black solid curve shows the normal
dwarf stars labeled with corresponding spectral types. Main-sequence stars and 
Giants with IR excess are plotted as blue, red plus symbol respectively. Red 
dotted line is the selection criterion from Section \ref{selection}. From
Figure \ref{k_22}, most of stars locate nearby the criterion line. The  
only one giant star locates $J-H>0.5$ in our sample. Most are excluded because 
of high 100 $\mu$m background level.

\begin{figure}[h!tb]
\begin{center}
\resizebox{15cm}{!}{\includegraphics{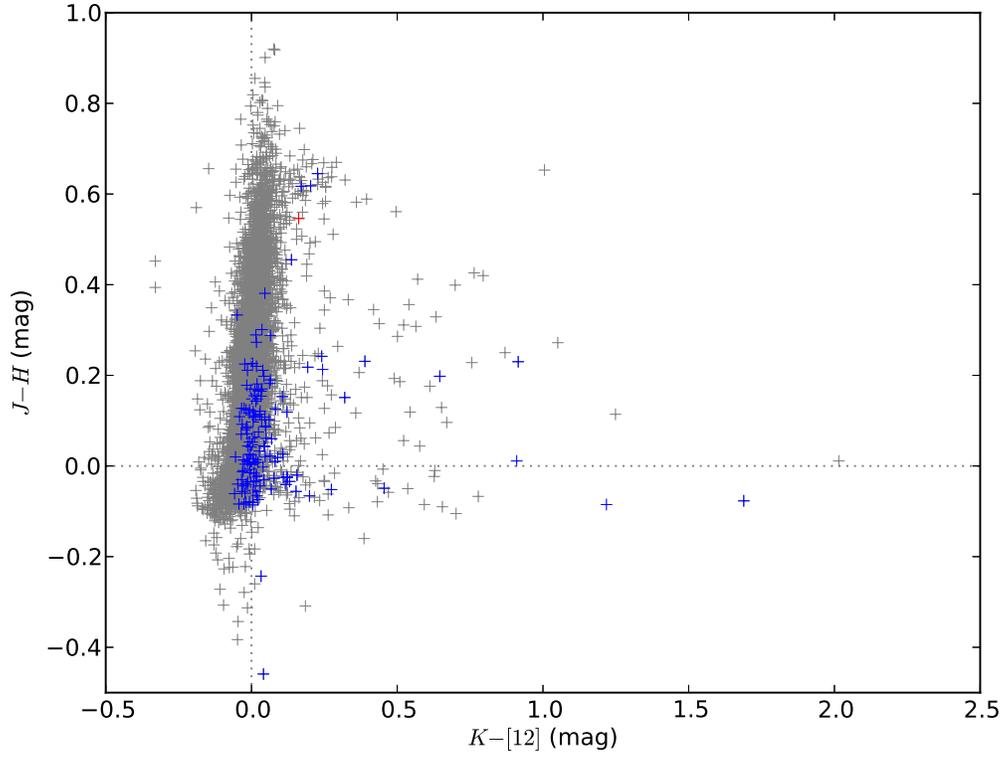}}
\end{center}
\caption{$K-[12]$ vs $J-H$ color-color diagram of 22 $\mu$m excess
    stars. Blue plus symbols and red plus symbols represent
    main-sequence stars and giants respectively. 
    Gray plus symbols mean all the
    matched sources from \emph{WISE} and \emph{Hipparcos}. \label{cc_diag}}
\end{figure}

Figure \ref{cc_diag} shows the $J-H$ vs $K-[12]$ color-color diagram
for the stars showing IR excess at 22 $\mu$m band. The blue plus signs 
represent the main-sequence stars and red plus signs for Giants. So as to be 
convenient for comparing, the x, y zero axes (gray dotted line) and all matched 
(\emph{WISE} and \emph{Hipparcos}) sources are plotted (gray plus signs). As 
shown in this figure, 
many main-sequence stars trend to have $K-[12]=0$. 
While the only one giant star trend to have excesses at 12 $\mu$m band. 
That is to say, comparing with the main-sequence stars, giants have larger 
luminosity at 12 $\mu$m band, and this implies that the dust around giants 
is hotter.

\begin{figure}[h!tb]
\begin{center}
\resizebox{15cm}{!}{\includegraphics{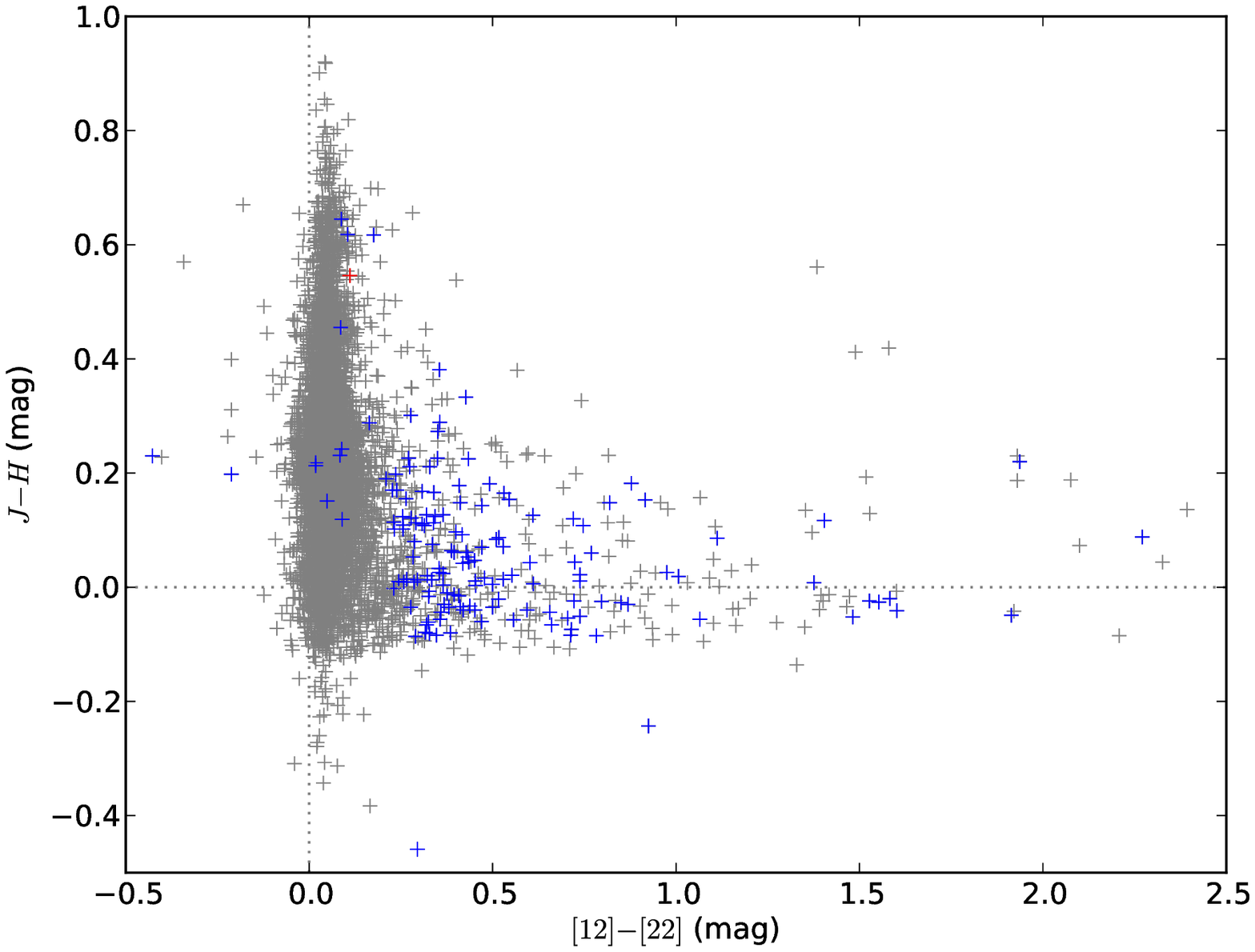}}
\end{center}
\caption{$W3-W4$ vs $J-H$ color-color diagram of 22 $\mu$m excess
    stars. Blue plus symbols and red plus symbols represent
    main-sequence stars and giants respectively. 
    Gray plus symbols mean all the
    matched sources from \emph{WISE} and \emph{Hipparcos}.
    Only a few stars locate neary the Y axis. \label{w3w4}}
\end{figure}

The $W3-W4$ vs $J-H$ color-color diagram is shown in Figure \ref{w3w4}. From 
Figure \ref{w3w4} we can see that very few stars locate nearby the Y axis. Those 
with $W3-W4\le0.1$ (left of or nearby the Y axis) are HIP 70386, HIP 22531, 
HIP 67953, HIP 30174, HIP 42753, HIP 12351, HIP 72138 and HIP 29888. Of the 8 
sources, they are all double or multiple stars except HIP 12531. That 
is to say, if the excess at 22 $\mu$m band is from the binary or the companion 
which can not be resolved by WISE, then $W3-W4$ should trend to be zero. While 
the trend can not be seen from Figure \ref{w3w4}. So we can conclude that the 
excess in most of our candidates is not from the binary or companion. The known 
binary and multiple systems have been noted in Table \ref{tab_main}. 

We have described in Section \ref{wise}, there is an overestimate at 4.6
$\mu$m band for brightness stars. We also found the effect in this work. When
we plot the $[3.4]-[4.6]$ vs $[3.4]-[12]$ (or $[3.4]-[22]$) diagram, we find 
that almost all of the giants and 1/4 of main-sequence stars located at
$[3.4]-[4.6]=0.5$ while most located at $[3.4]-[12]=0$, that is unreasonable! 
The phenomenon is similar to the bias mentioned in \citet{tisserand2012} and 
the bias found in this paper is about 0.5 mag. WISE team also reported this 
effect but gave no explain.

\subsection{The Dust Fraction  \label{fd}}
In order to characterize the amount of dust, the ratio of integrated
infrared excess luminosity to bolometric one of a star,
$f_{\rm{d}}=L_{\rm{IR}}/L_*=F_{\rm{IR}}/F_*$ \citep{moor2006, Carpenter2009} is 
introduced. In this
section, we will introduce two method to estimate the dust fraction 
$f_{\rm{d}}$ for all the candidates. 

First, because we don't have far-infrared 
fluxes for all these stars because of no longer bands in\emph{WISE}. So we 
assumed $\nu L_{\nu}$ as the total infrared luminosity $L_{\rm{IR}}$ 
\citep{chen2005a, wu2012}. Considering the bolometric luminosities of stars, 
we integrated the whole flux with different band. 
Then the dust fraction for each candidate can be estimated which is listed in 
Table \ref{tab_main}. 

\begin{figure}[h!tb]
\begin{center}
\resizebox{15cm}{!}{\includegraphics{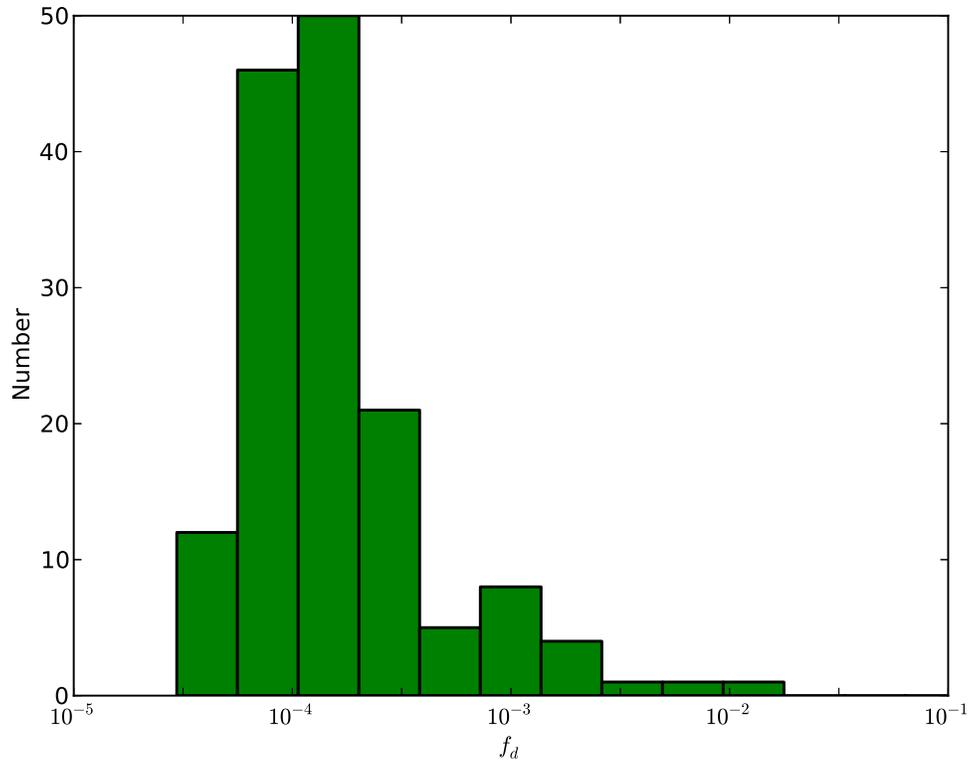}}
\end{center}
\caption{Histgram for the fractional luminosity $f_d$. 
    \label{fd_hist}}
\end{figure}

The second method is from \citet{beichman2005}. 
\begin{equation}
    \frac{L_{dust}}{L_*}(minimum)=10^{-5}(\frac{5600 K}{T_*})^3
    \frac{F_{70,dust}}{F_{70,*}}
\end{equation}
The fractional luminosity derived by \citet{beichman2005} depends on the dust 
temperature. We set the emission peak at 22 $\mu$m, which means 
$T_{dust}\approx150$ K. 
Then the minimum fractional luminosity can be calculated. 
The results from the two methods mainly range from $10^{-5}$ even to $10^{-3}$. 
The histogram of dust fraction for all candidates is plotted in Figure 
\ref{fd_hist}.

%
\citet{beichman2005} point that the fractional luminosity can be
used as an age indicator. But there is no yet final
conclusion. \citet{zuckerman2004} hypothesesed that stars with
$f_d>10^{-3}$ are younger than 100 Myr and \citet{beichman2005} also
suggested a decline in the fraction of stars with excess IR emission
with time. On the contrary, \citet{decin2003} claimed the existence of
high $f_d$ disks around older stars and they also found that the
fractional luminosity show a large spread ($10^{-6} - 10^{-3}$) at almost
any age. Although we don't know yet which point of the two is
supported in this paper, we can obtain one another conclusion.
Obviously it can be seen in Figure
\ref{fd_hist} that the giants have higher $f_d$ than main-sequence
stars. It can be explained as the stars at the late evolutionary
stages have more luminosity at mid-infrared band, which is also
consistent with the characteristics of giants.

If some older stars with higher $f_d$ in our galaxy, they are really
rare systems up to date \citep{wu2012}. Although \citet{rieke2005}
suggested that the dust would be more plentiful in the late stages of
planet formation from planetesimals collisions and cometary activity,
this can not fully explain the high fractional luminosity phenomena of
the old stars. The most plausible explanation for the presence of debris
disk with high fractional luminosity at ages well above Gyr is delay
onset of collisional cascades by late planet formation further away
from the star \citep{dominik2003}. However whether such a mechanism
can also explain very old stars at ages of 10Gyrs is still
questionable. The bottom line is to confirm the ages of these stars
with the high fractional luminosity.

\subsection{Images and Spectral Energy Distribution}
We present all the optical images and SEDs for our sample. The figures
are shown in appendix \ref{appendixc}. The images are from DSS (Digitized Sky
Survey) and the positions are also marked on the optical images as
red circle with a radius of $6^{\prime\prime}$. The SEDs cover all the
wavelength range from the optical to the mid-IR bands, including the
available photometry data from \emph{Hipparcos} (B, V and I), 2MASS
(J, H and $\rm{K_s}$), and \emph{WISE} (W3 and W4). All these SEDs 
are fitted by the blackbody formula. Our candidates is very bright, to avoid 
the influence of saturation, so the [3.4] and [4.6] band do not be used for 
SEDs fitting. The SEDs provided in this paper can be used to test our 
$K-[22]$ method of searching IR excess.

\subsection{Candidates with Planets and Debris Disks Candidates \label{planets}}
\citet{bryden2009} have searched for debris disks and planets using
\emph{Spitzer}'s MIPS far-IR camera, but did not find the correlation
between planets and dust around. We 
cross-matched our catalog with known IR-excess Hipparcos stars hosting planets
\citep{maldonado2012}. It turns out that none was found. Similar with 
\citet{kennedy2012}, those matched may all be excluded from our analysis 
because their higher background level or higher photometric error.
We also cross-matched our catalog with \emph{Kepler} 
planet candidates \citep{kepler2012}, none 
was found. This is because our samples are brighter than the \emph{Kepler} 
candidates. We still compared our candidates with debris disks candidates in 
\citet{rhee2007}, which studied the debris disks from the \emph{IRAS} and 
\emph{Hipparcos} catalog, and 27 sources are matched, they are all 
noted in the total catalog (Table \ref{tab_main}). By comparing with previous 
conclusion, there are 38 known IR excess stars or deberis disk candidates. They 
are all noted in Table \ref{tab_main}.
Those not be matched may contain the new debris disks candidates which can 
not be found by \emph{IRAS}.

By doing this, we want to know that if we can confirm the existence of the
extra-solar planet around an IR-excess star, we could attribute the IR excess 
to their asteroid belt. But we don't found known candidates with planets 
from the 141 IR excess stars. This can be explained by two points. One is that 
the excess of the 141 stars is from only dust, the other is that some unknown 
candidates with planets included have not been found.

\subsection{Fraction of Main-Sequence Stars with IR Excess}
In our sample, we count the number of stars with IR excess for
main-sequence stars.
Using the threshold mentioned in Section
\ref{hipparcos}, the cross-matched catalog contains about 7,624 
sources whose spectral types are either main-sequence or giants.
Filtering the giant stars with $M_v>6.0(B-V)-2.0$ and dropping those with  
high IRAS 100 $\mu$m background level, we found about
2,649 main-sequence sources without background contamination, which yields the 
fraction of main-sequence stars, about $140/2649\sim5.30\%$, while the 
fraction of FGK main-sequence stars is $42/1897\sim2.21\%$. Noting that the 
number here is less than the $9\% \sim 17\%$ derived by 
\citet{hovhannisyan2009}, $10\%$ concluded by \citet{meyer2008}. This is mainly 
because we set the definition of IR excess with 4$\sigma$. When 3$\sigma$ 
is used, then the detection rate of FGK main-sequence is 
$130/1897\sim6.85\%$. Of which, 130 means the number of FGK main-sequence stars 
with 22 $\mu$m excess by using 3$\sigma$ definition. In addition, there are 
other reasons. Because we focus only on the very bright all-sky stars and 
exclude the stars with higher 
\emph{IRAS} 100 $\mu$m background. It means that almost all the stars locate in 
the galactic disk or star formation region are removed. According to the common 
sense, the IR excess stars locate in the low galactic latitude (like galactic 
disk or star formation region) are more than those locate in the high 
galactic latitude. Moreover, the originally selecting criterion of W4 
$S/N\ge20$ can lower the detection rate. 

Figure \ref{spt_hist} shows the distribution of main-sequence stars 
with IR excess, the Y axis is the detection rates of different spectral types. 
The corresponding error bars (1$\sigma$) are also drawn. The error bars are 
calculated by using the error of $K-[22]$, so the upper limit and the lower 
limit are not the same. Because K and M type stars with 22 $\mu$m excess are 
very few, we hardly see the error bar of K typr stars and the lower limit of 
M stars. From Figure \ref{spt_hist} we can see that B, A and M type stars have 
more IR excess than others. More detailed numbers can be seen in Table 
\ref{tab_detec}. 

\section{SUMMARY \label{summary}}
We present a catalog which includes 141 IR excess stars at
22$\mu$m.  All these stars are selected from \emph{WISE} All-sky data
cross-correlated with \emph{Hipparcos} Main Catalog. By examining
the WISE data for these selected candidates, we conclude that they all
show an IR excess at 22 $\mu$m (i.e., $K_s-[22]\ge0.26$ for 
$J-H\le0.1$, $K_s-[22]\ge0.21$ for $0.1<J-H\le0.3$, 
$K_s-[22]\ge0.22$ for $0.3<J-H\le0.5$ and $K_s-[22]\ge0.22$ for $J-H>0.5$). 
With
the help of \emph{Hipparcos} Main Catalog, we can classify them as
different type for detailed study according to IR excess production
mechanism, and the corresponding catalogs are given in appendix. In
this paper, we provide, for all the IR-excess candidates, the SEDs
from optical to mid-infrared and optical images, and we also give the
infrared dust fraction. 
Reading from
the color-color diagram, we can tell that most main sequence stars show IR 
excess at only 22 $\mu$m while giants at both 12 $\mu$m and 22 $\mu$m.

Generally speaking, the IR excess stars could be used to search for 
for exoplanets, so we cross-matched our catalog with known
IR excess stars having planets and \emph{Kepler} planet candidates, but none 
is matched. They are all filtered out by our criterion, the explanations are 
also given. 

Lastly, we count the number of stars showing IR excess for different spectral
type of main-sequence stars and also give the explanations to them.

The IR excess star catalog in this work can be used as the input for
many future study, e.g., searching extra-solar planets, searching stars
with debris disk and modeling the surrounding disk, studying the
mechanism of showing IR excess.

\acknowledgements
\section{Acknowledgements}
Ch-J, Wu thanks Li-Jun Gou, Lian Yang and Jun-Li Cao for warmhearted
help and Xu Zhou for valuable discussion. we sincerely thank the anonymous 
referee whose suggestions greatly helped us improve this paper.
This project is supported by the 
China Ministry of Science and Technology under the State Key Development 
Program for Basic Research (2014CB845705, 2012CB821800), the National Natural 
Science Foundation
of China (Grant Nos. 11173030, 11225316, 11078017, 10833006, 10978014
and 10773014), the Key Laboratory of Optical Astronomy, the National
Astronomical Observatories, Chinese Academy of Sciences, the National
Science Foundation of China (Grants No. 11173006), the Ministry of
Science and Technology National Basic Science program (project 973)
under grant No. 2012CB821804, and the Fundamental Research Funds for
the Central Universities.

This publication makes use of data products from the Wide-field Infrared Survey
Explorer, which is a joint project of the University of California, Los
Angeles, and the Jet Propulsion Laboratory/California Institute of Technology,
funded by the National Aeronautics and Space Administration.

In this work, we made extensive use of the Hipparcos Catalogs (ESA 1997),
which is the primary result of the Hipparcos space astrometry mission,
undertaken by the European Space Agency, with the scientiﬁc aspects undertaken
by nearly two hundred scientists within the NDAC, FAST, TDAC and INCA
Consortia.

Facilities: 2.16m Telescope (NAOC), WISE, HIPPARCOS

\bibliographystyle{apj}
\bibliography{wu}

\begin{thebibliography}{61}
\expandafter\ifx\csname natexlab\endcsname\relax\def\natexlab#1{#1}\fi

\bibitem[{{Aumann} {et~al.}(1984){Aumann}, {Beichman}, {Gillett},
  {et~al.}}]{aumann84}
{Aumann}, H.~H., {Beichman}, C.~A., {Gillett}, F.~C., {et~al.} 1984, \apjl,
  278, L23

\bibitem[{{Avenhaus} {et~al.}(2012){Avenhaus}, {Schmid}, \&
  {Meyer}}]{avenhaus2012}
{Avenhaus}, H., {Schmid}, H.~M., \& {Meyer}, M.~R. 2012, \aap, 548, A105

\bibitem[{{Batalha} {et~al.}(2012){Batalha}, {Rowe}, {Bryson},
  {et~al.}}]{kepler2012}
{Batalha}, N.~M., {Rowe}, J.~F., {Bryson}, S.~T., {et~al.} 2012, ArXiv
  e-prints: 1202.5852

\bibitem[{{Beichman} {et~al.}(2005){Beichman}, {Bryden}, {Rieke},
  {et~al.}}]{beichman2005}
{Beichman}, C.~A., {Bryden}, G., {Rieke}, G.~H., {et~al.} 2005, \apj, 622, 1160

\bibitem[{{Bertout} {et~al.}(1999){Bertout}, {Robichon}, \&
  {Arenou}}]{bertout1999}
{Bertout}, C., {Robichon}, N., \& {Arenou}, F. 1999, \aap, 352, 574

\bibitem[{{Bryden} {et~al.}(2009){Bryden}, {Beichman}, {Carpenter},
  {et~al.}}]{bryden2009}
{Bryden}, G., {Beichman}, C.~A., {Carpenter}, J.~M., {et~al.} 2009, \apj, 705,
  1226

\bibitem[{{Carpenter} {et~al.}(2009){Carpenter}, {Bouwman}, {Mamajek}, {Meyer},
  {Hillenbrand}, {Backman}, {Henning}, {Hines}, {Hollenbach}, {Kim},
  {Moro-Martin}, {Pascucci}, {Silverstone}, {Stauffer}, \&
  {Wolf}}]{Carpenter2009}
{Carpenter}, J.~M., {Bouwman}, J., {Mamajek}, E.~E., {Meyer}, M.~R.,
  {Hillenbrand}, L.~A., {Backman}, D.~E., {Henning}, T., {Hines}, D.~C.,
  {Hollenbach}, D., {Kim}, J.~S., {Moro-Martin}, A., {Pascucci}, I.,
  {Silverstone}, M.~D., {Stauffer}, J.~R., \& {Wolf}, S. 2009, \apjs, 181, 197

\bibitem[{{Chen} {et~al.}(2005{\natexlab{a}}){Chen}, {Jura}, {Gordon}, \&
  {Blaylock}}]{chen2005a}
{Chen}, C.~H., {Jura}, M., {Gordon}, K.~D., \& {Blaylock}, M.
  2005{\natexlab{a}}, \apj, 623, 493

\bibitem[{{Chen} {et~al.}(2005{\natexlab{b}}){Chen}, {Patten}, {Werner},
  {et~al.}}]{chen2005b}
{Chen}, C.~H., {Patten}, B.~M., {Werner}, M.~W., {et~al.} 2005{\natexlab{b}},
  \apj, 634, 1372

\bibitem[{{Ciliegi} {et~al.}(2003){Ciliegi}, {Zamorani}, {Hasinger},
  {et~al.}}]{ciliegi2003}
{Ciliegi}, P., {Zamorani}, G., {Hasinger}, G., {et~al.} 2003, \aap, 398, 901

\bibitem[{{Dame} {et~al.}(2001){Dame}, {Hartmann}, \& {Thaddeus}}]{dame2001}
{Dame}, T.~M., {Hartmann}, D., \& {Thaddeus}, P. 2001, \apj, 547, 792

\bibitem[{{Debes} {et~al.}(2011){Debes}, {Hoard}, {Wachter},
  {et~al.}}]{debes2011}
{Debes}, J.~H., {Hoard}, D.~W., {Wachter}, S., {et~al.} 2011, \apjs, 197, 38

\bibitem[{{Decin} {et~al.}(2003){Decin}, {Dominik}, {Waters}, \&
  {Waelkens}}]{decin2003}
{Decin}, G., {Dominik}, C., {Waters}, L.~B.~F.~M., \& {Waelkens}, C. 2003,
  \apj, 598, 636

\bibitem[{{Dominik} \& {Decin}(2003)}]{dominik2003}
{Dominik}, C. \& {Decin}, G. 2003, \apj, 598, 626

\bibitem[{{G{\'a}sp{\'a}r} {et~al.}(2013){G{\'a}sp{\'a}r}, {Rieke}, \&
  {Balog}}]{andras2013}
{G{\'a}sp{\'a}r}, A., {Rieke}, G.~H., \& {Balog}, Z. 2013, \apj, 768, 25

\bibitem[{{Gorlova} {et~al.}(2004){Gorlova}, {Padgett}, {Rieke},
  {et~al.}}]{gorlova2004}
{Gorlova}, N., {Padgett}, D.~L., {Rieke}, G.~H., {et~al.} 2004, \apjs, 154, 448

\bibitem[{{Gorlova} {et~al.}(2006){Gorlova}, {Rieke}, {Muzerolle},
  {et~al.}}]{gorlova2006}
{Gorlova}, N., {Rieke}, G.~H., {Muzerolle}, J., {et~al.} 2006, \apj, 649, 1028

\bibitem[{{Hoard} {et~al.}(2007){Hoard}, {Wachter}, {Sturch},
  {et~al.}}]{hoard2007}
{Hoard}, D.~W., {Wachter}, S., {Sturch}, L.~K., {et~al.} 2007, \aj, 134, 26

\bibitem[{{Hovhannisyan} {et~al.}(2009){Hovhannisyan}, {Mickaelian}, {Weedman},
  {et~al.}}]{hovhannisyan2009}
{Hovhannisyan}, L.~R., {Mickaelian}, A.~M., {Weedman}, D.~W., {et~al.} 2009,
  \aj, 138, 251

\bibitem[{{Ivison} {et~al.}(2007){Ivison}, {Greve}, {Dunlop},
  {et~al.}}]{ivison2007}
{Ivison}, R.~J., {Greve}, T.~R., {Dunlop}, J.~S., {et~al.} 2007, \mnras, 380,
  199

\bibitem[{{Kennedy} \& {Wyatt}(2012)}]{kennedy2012}
{Kennedy}, G.~M. \& {Wyatt}, M.~C. 2012, \mnras, 426, 91

\bibitem[{{Kenyon} {et~al.}(1994){Kenyon}, {Dobrzycka}, \&
  {Hartmann}}]{kenyon1994}
{Kenyon}, S.~J., {Dobrzycka}, D., \& {Hartmann}, L. 1994, \aj, 108, 1872

\bibitem[{{Kim} {et~al.}(2005){Kim}, {Hines}, {Backman}, {et~al.}}]{kim2005}
{Kim}, J.~S., {Hines}, D.~C., {Backman}, D.~E., {et~al.} 2005, \apj, 632, 659

\bibitem[{{Knude} \& {Hog}(1998)}]{knude1998}
{Knude}, J. \& {Hog}, E. 1998, \aap, 338, 897

\bibitem[{{Koerner} {et~al.}(2010){Koerner}, {Kim}, {Trilling},
  {et~al.}}]{koerner2010}
{Koerner}, D.~W., {Kim}, S., {Trilling}, D.~E., {et~al.} 2010, \apjl, 710, L26

\bibitem[{{Lagrange} {et~al.}(2000){Lagrange}, {Backman}, \&
  {Artymowicz}}]{lagrange2000}
{Lagrange}, A.-M., {Backman}, D.~E., \& {Artymowicz}, P. 2000, Protostars and
  Planets IV, 639

\bibitem[{{Lawler} \& {Gladman}(2012)}]{lawler2012}
{Lawler}, S.~M. \& {Gladman}, B. 2012, \apj, 752, 53

\bibitem[{{Low} {et~al.}(2005){Low}, {Smith}, {Werner}, {et~al.}}]{low2005}
{Low}, F.~J., {Smith}, P.~S., {Werner}, M., {et~al.} 2005, \apj, 631, 1170

\bibitem[{{Mainieri} {et~al.}(2008){Mainieri}, {Kellermann}, {Fomalont},
  {et~al.}}]{mainieri2008}
{Mainieri}, V., {Kellermann}, K.~I., {Fomalont}, E.~B., {et~al.} 2008, \apjs,
  179, 95

\bibitem[{{Maldonado} {et~al.}(2012){Maldonado}, {Eiroa}, {Villaver},
  {et~al.}}]{maldonado2012}
{Maldonado}, J., {Eiroa}, C., {Villaver}, E., {et~al.} 2012, \aap, 541, A40

\bibitem[{{Mamajek}(2008)}]{mamajek2008}
{Mamajek}, E.~E. 2008, Astronomische Nachrichten, 329, 10

\bibitem[{{Meyer} {et~al.}(2008){Meyer}, {Carpenter}, {Mamajek},
  {et~al.}}]{meyer2008}
{Meyer}, M.~R., {Carpenter}, J.~M., {Mamajek}, E.~E., {et~al.} 2008, \apjl,
  673, L181

\bibitem[{{Meyer} {et~al.}(2004){Meyer}, {Hillenbrand}, {Backman},
  {et~al.}}]{meyer2004}
{Meyer}, M.~R., {Hillenbrand}, L.~A., {Backman}, D.~E., {et~al.} 2004, \apjs,
  154, 422

\bibitem[{{Mizusawa} {et~al.}(2012){Mizusawa}, {Rebull}, {Stauffer}, {Bryden},
  {Meyer}, \& {Song}}]{Mizusawa2012}
{Mizusawa}, T.~F., {Rebull}, L.~M., {Stauffer}, J.~R., {Bryden}, G., {Meyer},
  M., \& {Song}, I. 2012, \aj, 144, 135

\bibitem[{{Mo{\'o}r} {et~al.}(2006){Mo{\'o}r}, {{\'A}brah{\'a}m}, {Derekas},
  {et~al.}}]{moor2006}
{Mo{\'o}r}, A., {{\'A}brah{\'a}m}, P., {Derekas}, A., {et~al.} 2006, \apj, 644,
  525

\bibitem[{{Morales} {et~al.}(2012){Morales}, {Padgett}, {Bryden}, {Werner}, \&
  {Furlan}}]{morales2012}
{Morales}, F.~Y., {Padgett}, D.~L., {Bryden}, G., {Werner}, M.~W., \& {Furlan},
  E. 2012, \apj, 757, 7

\bibitem[{{Morales} {et~al.}(2009){Morales}, {Werner}, {Bryden}, {Plavchan},
  {Stapelfeldt}, {Rieke}, {Su}, {Beichman}, {Chen}, {Grogan}, {Kenyon},
  {Moro-Martin}, \& {Wolf}}]{Morales2009}
{Morales}, F.~Y., {Werner}, M.~W., {Bryden}, G., {Plavchan}, P., {Stapelfeldt},
  K.~R., {Rieke}, G.~H., {Su}, K.~Y.~L., {Beichman}, C.~A., {Chen}, C.~H.,
  {Grogan}, K., {Kenyon}, S.~J., {Moro-Martin}, A., \& {Wolf}, S. 2009, \apj,
  699, 1067

\bibitem[{{Perryman} {et~al.}(1995){Perryman}, {Lindegren}, {Kovalevsky},
  {et~al.}}]{perryman1995}
{Perryman}, M.~A.~C., {Lindegren}, L., {Kovalevsky}, J., {et~al.} 1995, \aap,
  304, 69

\bibitem[{{Perryman} {et~al.}(1997){Perryman}, {Lindegren}, {Kovalevsky},
  {et~al.}}]{perryman1997}
---. 1997, \aap, 323, L49

\bibitem[{{Rebull} {et~al.}(2008){Rebull}, {Stapelfeldt}, {Werner}, {Mannings},
  {Chen}, {Stauffer}, {Smith}, {Song}, {Hines}, \& {Low}}]{Rebull2008}
{Rebull}, L.~M., {Stapelfeldt}, K.~R., {Werner}, M.~W., {Mannings}, V.~G.,
  {Chen}, C., {Stauffer}, J.~R., {Smith}, P.~S., {Song}, I., {Hines}, D., \&
  {Low}, F.~J. 2008, \apj, 681, 1484

\bibitem[{{Rhee} {et~al.}(2007){Rhee}, {Song}, {Zuckerman}, \&
  {McElwain}}]{rhee2007}
{Rhee}, J.~H., {Song}, I., {Zuckerman}, B., \& {McElwain}, M. 2007, \apj, 660,
  1556

\bibitem[{{Ribas} {et~al.}(2012){Ribas}, {Mer{\'{\i}}n}, {Ardila}, \&
  {Bouy}}]{ribas2012}
{Ribas}, {\'A}., {Mer{\'{\i}}n}, B., {Ardila}, D.~R., \& {Bouy}, H. 2012, \aap,
  541, A38

\bibitem[{{Richter}(1975)}]{richter1975}
{Richter}, G.~A. 1975, Astronomische Nachrichten, 296, 65

\bibitem[{{Rieke} {et~al.}(2005){Rieke}, {Su}, {Stansberry},
  {et~al.}}]{rieke2005}
{Rieke}, G.~H., {Su}, K.~Y.~L., {Stansberry}, J.~A., {et~al.} 2005, \apj, 620,
  1010

\bibitem[{{Rieke} {et~al.}(2004){Rieke}, {Young}, {Engelbracht},
  {et~al.}}]{rieke2004}
{Rieke}, G.~H., {Young}, E.~T., {Engelbracht}, C.~W., {et~al.} 2004, \apjs,
  154, 25

\bibitem[{{Smith} {et~al.}(2011){Smith}, {Dunne}, {Maddox},
  {et~al.}}]{smith2011}
{Smith}, D.~J.~B., {Dunne}, L., {Maddox}, S.~J., {et~al.} 2011, \mnras, 416,
  857

\bibitem[{{Stauffer} {et~al.}(2005){Stauffer}, {Rebull}, {Carpenter},
  {et~al.}}]{stauffer2005}
{Stauffer}, J.~R., {Rebull}, L.~M., {Carpenter}, J., {et~al.} 2005, \aj, 130,
  1834

\bibitem[{{Su} {et~al.}(2006){Su}, {Rieke}, {Stansberry}, {et~al.}}]{su2006}
{Su}, K.~Y.~L., {Rieke}, G.~H., {Stansberry}, J.~A., {et~al.} 2006, \apj, 653,
  675

\bibitem[{{Sutherland} \& {Saunders}(1992)}]{sutherland1992}
{Sutherland}, W. \& {Saunders}, W. 1992, \mnras, 259, 413

\bibitem[{{Thompson}(1982)}]{thompson1982}
{Thompson}, R.~I. 1982, \apj, 257, 171

\bibitem[{{Tisserand}(2012)}]{tisserand2012}
{Tisserand}, P. 2012, \aap, 539, A51

\bibitem[{{Touhami} {et~al.}(2011){Touhami}, {Gies}, \&
  {Schaefer}}]{Touhami2011}
{Touhami}, Y., {Gies}, D.~R., \& {Schaefer}, G.~H. 2011, \apj, 729, 17

\bibitem[{{Wachter} {et~al.}(2003){Wachter}, {Hoard}, {Hansen},
  {et~al.}}]{wachter2003}
{Wachter}, S., {Hoard}, D.~W., {Hansen}, K.~H., {et~al.} 2003, \apj, 586, 1356

\bibitem[{{Wellhouse} {et~al.}(2005){Wellhouse}, {Hoard}, {Howell},
  {et~al.}}]{wellhouse2005}
{Wellhouse}, J.~W., {Hoard}, D.~W., {Howell}, S.~B., {et~al.} 2005, \pasp, 117,
  1378

\bibitem[{{Werner} {et~al.}(2004){Werner}, {Uchida}, {Sellgren},
  {et~al.}}]{werner2004}
{Werner}, M.~W., {Uchida}, K.~I., {Sellgren}, K., {et~al.} 2004, \apjs, 154,
  309

\bibitem[{{Wright} {et~al.}(2010){Wright}, {Eisenhardt}, {Mainzer},
  {et~al.}}]{wright2010}
{Wright}, E.~L., {Eisenhardt}, P.~R.~M., {Mainzer}, A.~K., {et~al.} 2010, \aj,
  140, 1868

\bibitem[{{Wu} {et~al.}(2012){Wu}, {Wu}, {Cao}, {et~al.}}]{wu2012}
{Wu}, H., {Wu}, C.-J., {Cao}, C., {et~al.} 2012, RAA, 12, 513

\bibitem[{{Young} {et~al.}(2004){Young}, {Lada}, {Teixeira},
  {et~al.}}]{young2004}
{Young}, E.~T., {Lada}, C.~J., {Teixeira}, P., {et~al.} 2004, \apjs, 154, 428

\bibitem[{{Zuckerman}(2001)}]{zuckerman2001}
{Zuckerman}, B. 2001, \araa, 39, 549

\bibitem[{{Zuckerman} {et~al.}(2011){Zuckerman}, {Rhee}, {Song}, \&
  {Bessell}}]{Zuckerman2011}
{Zuckerman}, B., {Rhee}, J.~H., {Song}, I., \& {Bessell}, M.~S. 2011, \apj,
  732, 61

\bibitem[{{Zuckerman} \& {Song}(2004)}]{zuckerman2004}
{Zuckerman}, B. \& {Song}, I. 2004, \araa, 42, 685

\end{thebibliography}

\clearpage

\clearpage


\clearpage

\clearpage

\clearpage

\clearpage

\clearpage

\clearpage

\clearpage

\clearpage

\clearpage

\clearpage

\clearpage

\clearpage

\clearpage


\clearpage
\begin{deluxetable}{lccccrclc}
\tabletypesize{\footnotesize}
\rotate
\tablecolumns{9}
\tablewidth{0pc}
\tablecaption{Log of observation and spectral type for 6 selected stars showing 
excess at 22 $\mu$m band \label{obs_stars}}
\tablehead{
\colhead{Name} & \colhead{RA} & \colhead{DEC} & \colhead{Instrument} 
               & \colhead{Slit} & \colhead{Plx} & \colhead{SpType} 
               & \colhead{SpType} & \colhead{Date of Obs.} \\
               & \colhead{(J2000)} & \colhead{(J2000)} & 
               & \colhead{(arcsec)} & \colhead{(mas)} 
               & \colhead{(2.16m Telescope)} & \colhead{($Hipparcos$)} &
}
\startdata
HIP 17681 & 03 47 16.10 & +44 04 25.7 & BFOSC Grism\#7  & 1.8  &  8.13 & B9 & B9 & (01.21.2012) \\
HIP 20063 & 04 18 08.09 & +42 08 28.5 & BFOSC Grism\#7  & 1.8  &  8.39 & B9 & B9V & (01.21.2012) \\
HIP 20713 & 04 26 20.67 & +15 37 06.0 & BFOSC Grism\#7  & 1.8  & 20.86 & F0 & F0V... & (01.21.2012) \\
HIP 21038 & 04 30 38.40 & +32 27 28.1 & BFOSC Grism\#7  & 1.8  & 10.63 & B9 & B9.5Vn & (01.21.2012) \\
HIP 25453 & 05 26 38.82 & +06 52 07.5 & BFOSC Grism\#7  & 1.8  & 10.68 & A0 & A0Vn & (01.21.2012) \\
HIP 26062 & 05 33 30.75 & +24 37 44.1 & BFOSC Grism\#7  & 1.8  & 10.0  & B8 & B8 & (01.21.2012) \\
\enddata
\end{deluxetable}

\clearpage
\begin{deluxetable}{llll}
\tabletypesize{\footnotesize}
\tablecolumns{4}
\tablewidth{0pc}
\tablecaption{22 $\mu$m Stars Excess Catalog Format \label{tab_sum}}
\tablehead{
    \colhead{Column} & \colhead{Name} & \colhead{Units} & \colhead{Description} 
}
\startdata
1  & HIP      & --- & Name of stars in the recommended format for Hipparcos stars \\ 
2  & RAdeg    & deg & Right Ascension 2000 (degrees)  \\
3  & DEdeg    & deg & Declination 2000 (degrees) \\
4  & B        & mag & Magnitude in Johnson B \\
5  &          & mag & Error of Magnitude in Johnson B \\
6  & V        & mag & Magnitude in Johnson V \\
7  &          & mag & Error of Magnitude in Johnson V \\
8  & I        & mag & Magnitude in Johnson I \\
9  &          & mag & Error of Magnitude in Johnson I \\
10 & J        & mag & 2MASS J-band magnitude \\
11 &          & mag & Error of 2MASS J-band magnitude \\
12 & H        & mag & 2MASS H-band magnitude \\
13 &          & mag & Error of 2MASS H-band magnitude \\
14 & $\rm{K_s}$        & mag & 2MASS K-band magnitude \\
15 &          & mag & Error of 2MASS K-band magnitude \\
16 & [3.4]    & mag & W1 magnitude of WISE \\
17 &          & mag & Error of W1 magnitude of WISE \\
18 & [4.6]    & mag & W2 magnitude of WISE \\ 
19 &          & mag & Error of W2 magnitude of WISE \\ 
20 & [12]     & mag & W3 magnitude of WISE \\
21 &          & mag & Error of W3 magnitude of WISE \\
22 & [22]     & mag & W4 magnitude of WISE \\
23 &          & mag & Error of W4 magnitude of WISE \\
24 & $K_s-[22]$ & mag & Criterion of searching for 22 $\mu$m excess stars \\
25 &          & mag & Error of $K_s-[22]$ \\
26 & $f_d$    & --- & Fractional Luminosity \\ 
27 & SpType   & --- & Spectral Type \\
28 & Note     & --- & Notes for stars \\
29 & Refs     & --- & Reference \\
\enddata
\end{deluxetable}

\section{Table of IR excess stars: The 140 Main-Sequence stars and 1 Giants}
\clearpage
\begin{deluxetable}{rrrcllllllllllllcl}
\tabletypesize{\scriptsize}
\rotate
\tablecolumns{16}
\tablewidth{0pc}
\tablecaption{Stars with 22 $\mu$m Excess \label{tab_main}}
\tablehead{
    \colhead{HIP} & \colhead{RAdeg} & \colhead{DEdeg} & \colhead{...} & 
    \multicolumn{2}{c}{$K_s$} & \colhead{...} &
     \multicolumn{2}{c}{$[12]$} & 
    \multicolumn{2}{c}{$[22]$} & \multicolumn{2}{c}{$K_s-[22]$} & 
    \colhead{$f_d$} & \colhead{SpType} & \colhead{Note} & 
    \colhead{Refs}\\
                    & \colhead{(J2000)} & \colhead{(J2000)} & \colhead{...} & 
    \multicolumn{2}{c}{(mag)} & \colhead{...} & 
    \multicolumn{2}{c}{(mag)} & 
    \multicolumn{2}{c}{(mag)} & \multicolumn{2}{c}{(mag)} & & & &
}
\startdata
301    & 0.93488485 & -17.33597   & \nodata &  4.56 & 0.02  & \nodata & 4.52   & 0.01  & 4.23  & 0.02  & 0.34  & 0.03  & 3.47e-05& B9      & &\\      
560    & 1.7084357  & -23.107426  & \nodata &  5.24 & 0.02  & \nodata & 5.23   & 0.01  & 4.51  & 0.02  & 0.73  & 0.03  & 3.61e-04& F2       & &\citet{Rebull2008}\\
682    & 2.1070592  & 6.6168075   & \nodata &  6.12 & 0.02  & \nodata & 6.1    & 0.01  & 5.75  & 0.04  & 0.37  & 0.05  & 2.84e-04& G2        & Disk &\citet{moor2006}\\
813    & 2.5091026  & 11.145809   & \nodata &  5.7  & 0.02  & \nodata & 5.62   & 0.01  & 5.38  & 0.04  & 0.33  & 0.04  & 3.39e-05& B9        & D or M &Simbad\\
2496   & 7.9197493  & -1.7936332  & \nodata &  6.89 & 0.02  & \nodata & 6.78   & 0.02  & 5.8   & 0.04  & 1.08  & 0.04  & 3.25e-04& A0         & C&\\
4366   & 13.993843  & 27.209354   & \nodata &  5.75 & 0.02  & \nodata & 5.75   & 0.01  & 5.29  & 0.03  & 0.45  & 0.04  & 1.17e-04& A5       & &\\
6507   & 20.895601  & -24.352776  & \nodata &  6.11 & 0.02  & \nodata & 6.05   & 0.01  & 5.82  & 0.04  & 0.29  & 0.04  & 1.15e-04& A9        & D or M&Simbad\\
6679   & 21.41935   & 2.972085    & \nodata &  5.9  & 0.02  & \nodata & 5.92   & 0.01  & 5.6   & 0.03  & 0.29  & 0.04  & 1.18e-04& F0         & C; in Double system&\\
7345   & 23.657173  & -15.6763525 & \nodata &  5.46 & 0.02  & \nodata & 5.34   & 0.01  & 3.74  & 0.02  & 1.72  & 0.03  & 1.54e-03& A1        & &\citet{rhee2007} \\
7805   & 25.099968  & -60.99904   & \nodata &  6.63 & 0.02  & \nodata & 6.6    & 0.01  & 6.05  & 0.04  & 0.58  & 0.04  & 2.56e-04& F2     & &\citet{rhee2007} \\
7943   & 25.514395  & 35.245766   & \nodata &  5.78 & 0.02  & \nodata & 5.75   & 0.01  & 5.04  & 0.03  & 0.74  & 0.03  & 8.77e-05& B9     & &\\
8122   & 26.09486   & 32.51604    & \nodata &  6.17 & 0.02  & \nodata & 6.18   & 0.01  & 5.66  & 0.04  & 0.51  & 0.04  & 1.16e-04& A3        & & \citet{rhee2007} \\
9570   & 30.74153   & 33.284153   & \nodata &  5.33 & 0.02  & \nodata & 5.34   & 0.01  & 5.06  & 0.03  & 0.27  & 0.03  & 6.04e-05& A2        & &\citet{rhee2007} \\
10320  & 33.226917  & -30.723843  & \nodata &  5.21 & 0.02  & \nodata & 5.08   & 0.01  & 4.66  & 0.03  & 0.54  & 0.03  & 9.37e-05& A0        & &\\
10670  & 34.328484  & 33.84732    & \nodata &  3.96 & 0.03  & \nodata & 3.98   & 0.01  & 3.51  & 0.02  & 0.45  & 0.03  & 8.20e-05& A1      & &\citet{rhee2007} \\
11360  & 36.567474  & 6.292674    & \nodata &  5.82 & 0.02  & \nodata & 5.79   & 0.01  & 5.26  & 0.03  & 0.56  & 0.04  & 2.48e-04& F2         & C; &\citet{rhee2007} \\
11477  & 37.007042  & -33.811054  & \nodata &  4.94 & 0.03  & \nodata & 4.98   & 0.01  & 4.56  & 0.02  & 0.38  & 0.03  & 7.80e-05& A2/A3     & &\\
11847  & 38.232403  & 37.33374    & \nodata &  6.55 & 0.02  & \nodata & 6.5    & 0.02  & 4.23  & 0.03  & 2.32  & 0.03  & 1.25e-02& F0         & &\citet{rhee2007} \\
12351  & 39.75498   & -58.18722   & \nodata &  5.76 & 0.02  & \nodata & 5.53   & 0.01  & 5.44  & 0.03  & 0.31  & 0.03  & 9.30e-04& M0       & &\\
12361  & 39.781216  & -52.93493   & \nodata &  5.92 & 0.03  & \nodata & 5.9    & 0.01  & 5.43  & 0.03  & 0.49  & 0.04  & 1.83e-04& F0/F2& C; &\citet{rhee2007} \\
13569  & 43.70644   & -33.524895  & \nodata &  6.22 & 0.02  & \nodata & 6.21   & 0.01  & 5.87  & 0.03  & 0.34  & 0.04  & 7.79e-05& A3        & &\\
15039  & 48.45911   & -38.809284  & \nodata &  6.43 & 0.02  & \nodata & 6.43   & 0.01  & 6.11  & 0.04  & 0.32  & 0.05  & 6.73e-05& A2/A3& &\\  
15933  & 51.303272  & -37.15268   & \nodata &  6.87 & 0.02  & \nodata & 6.83   & 0.02  & 6.59  & 0.05  & 0.27  & 0.05  & 7.27e-05& A4      &  & \\
16449  & 52.973385  & -25.614113  & \nodata &  6.1  & 0.02  & \nodata & 6.11   & 0.01  & 5.39  & 0.03  & 0.71  & 0.04  & 1.82e-04& A3     & &\citet{rhee2007} \\
16638  & 53.51848   & -41.33871   & \nodata &  6.67 & 0.02  & \nodata & 6.63   & 0.01  & 6.41  & 0.04  & 0.26  & 0.04  & 1.24e-04& F2        && \\
17395  & 55.890995  & -10.485612  & \nodata &  5.08 & 0.02  & \nodata & 5.08   & 0.01  & 4.66  & 0.03  & 0.42  & 0.03  & 1.08e-04& A5        & Disk &\citet{andras2013}\\
17681  & 56.817085  & 44.07382    & \nodata &  7.08 & 0.02  & \nodata & 6.97   & 0.01  & 6.17  & 0.04  & 0.91  & 0.05  & 1.29e-04& B9         & &\\
18187  & 58.36331   & -41.22287   & \nodata &  5.98 & 0.02  & \nodata & 5.95   & 0.01  & 5.64  & 0.03  & 0.33  & 0.04  & 1.81e-04& F6        & IR excess &\citet{Mizusawa2012}\\
18437  & 59.122307  & -38.96217   & \nodata &  6.86 & 0.02  & \nodata & 6.71   & 0.01  & 5.12  & 0.02  & 1.74  & 0.03  & 1.47e-03& A0        & &\citet{rhee2007} \\
18671  & 59.975647  & -54.161247  & \nodata &  6.78 & 0.02  & \nodata & 6.72   & 0.01  & 6.51  & 0.04  & 0.27  & 0.05  & 1.49e-04& F5        & &\\
20063  & 64.53372   & 42.141247   & \nodata &  6.35 & 0.02  & \nodata & 6.37   & 0.02  & 5.66  & 0.04  & 0.69  & 0.05  & 7.79e-05& B9        &  &\\
20713  & 66.586136  & 15.618346   & \nodata &  4.03 & 0.24  & \nodata & 3.78   & 0.01  & 3.77  & 0.03  & 0.26  & 0.24  & 1.09e-04& F0     &  &\\
20737  & 66.66073   & -28.951828  & \nodata &  6.74 & 0.02  & \nodata & 6.7    & 0.01  & 6.34  & 0.05  & 0.4   & 0.05  & 4.15e-04& K0        & &\\
21020  & 67.611206  & -43.212772  & \nodata &  7.31 & 0.02  & \nodata & 7.23   & 0.01  & 6.38  & 0.05  & 0.93  & 0.05  & 2.27e-04& A0        & &\\
21024  & 67.62306   & -43.410667  & \nodata &  7.22 & 0.02  & \nodata & 7.09   & 0.02  & 5.54  & 0.03  & 1.68  & 0.04  & 1.27e-03& A0        & &\\
21038  & 67.66001   & 32.457813   & \nodata &  6.2  & 0.02  & \nodata & 6.19   & 0.02  & 5.9   & 0.04  & 0.29  & 0.05  & 3.15e-05& B9     & Double?&\\
21618  & 69.6201    & -19.640846  & \nodata &  6.49 & 0.02  & \nodata & 6.48   & 0.01  & 6.19  & 0.04  & 0.3   & 0.05  & 1.15e-04& A9       & &\\
21765  & 70.12219   & -9.195834   & \nodata &  6.27 & 0.02  & \nodata & 6.07   & 0.01  & 5.96  & 0.05  & 0.31  & 0.05  & 9.15e-04& M0     & Double?&\\
22226  & 71.706276  & -26.302446  & \nodata &  6.89 & 0.02  & \nodata & 6.87   & 0.01  & 6.05  & 0.04  & 0.84  & 0.05  & 5.10e-04& F3        & &\citet{rhee2007} \\
22531  & 72.730774  & -53.46172   & \nodata &  4.8  & 0.02  & \nodata & 4.16   & 0.01  & 4.37  & 0.02  & 0.43  & 0.03  & 1.62e-04& F0    & in Double system&Simbad\\
24528  & 78.93295   & -22.894373  & \nodata &  6.43 & 0.03  & \nodata & 6.36   & 0.02  & 5.59  & 0.04  & 0.84  & 0.05  & 2.44e-04& A3        & &\citet{rhee2007} \\
24947  & 80.15841   & -39.754974  & \nodata &  6.14 & 0.02  & \nodata & 6.1    & 0.01  & 5.87  & 0.03  & 0.28  & 0.04  & 1.61e-04& F6        & Disk &\citet{Zuckerman2011}\\
25183  & 80.801575  & -31.748575  & \nodata &  6.41 & 0.02  & \nodata & 6.41   & 0.01  & 6.06  & 0.04  & 0.35  & 0.04  & 1.65e-04& F3        & Binary; IR excess identiﬁed &\citet{Mizusawa2012}\\
25453  & 81.661766  & 6.8687367   & \nodata &  6.44 & 0.02  & \nodata & 6.32   & 0.02  & 4.79  & 0.03  & 1.65  & 0.04  & 1.20e-03& A0       & &\\
25517  & 81.85334   & -26.584913  & \nodata &  6.7  & 0.04  & \nodata & 6.66   & 0.01  & 6.23  & 0.05  & 0.48  & 0.06  & 1.07e-04& A3  & C &\\
25608  & 82.063866  & -37.23093   & \nodata &  5.5  & 0.02  & \nodata & 5.52   & 0.01  & 5.22  & 0.03  & 0.28  & 0.03  & 5.10e-05& A0       & & \\
25998  & 83.23498   & -47.689034  & \nodata &  7.34 & 0.02  & \nodata & 7.31   & 0.01  & 5.94  & 0.03  & 1.4   & 0.04  & 8.86e-04& A3        & C &\\
26062  & 83.37814   & 24.628914   & \nodata &  6.82 & 0.02  & \nodata & 5.13   & 0.01  & 2.28  & 0.02  & 4.53  & 0.03  & 4.34e-01& B8         & &\\
26309  & 84.04282   & -28.708006  & \nodata &  5.86 & 0.02  & \nodata & 5.8    & 0.01  & 5.06  & 0.03  & 0.8   & 0.03  & 2.04e-04& A2   & &\\
26453  & 84.41505   & -28.626286  & \nodata &  6.28 & 0.02  & \nodata & 6.21   & 0.01  & 5.34  & 0.03  & 0.94  & 0.04  & 6.43e-04& F3        & C; &\citet{rhee2007} \\
26621  & 84.8769    & -40.68407   & \nodata &  7.1  & 0.02  & \nodata & 7.02   & 0.01  & 6.41  & 0.04  & 0.69  & 0.05  & 6.28e-05& B8        & &\\
26796  & 85.362175  & -33.40071   & \nodata &  6.43 & 0.03  & \nodata & 6.4    & 0.02  & 5.74  & 0.03  & 0.69  & 0.04  & 7.78e-05& B9      & Excess at 24 $\mu$m &\citet{Morales2009}\\
26966  & 85.84025   & -18.557444  & \nodata &  5.78 & 0.03  & \nodata & 5.63   & 0.02  & 4.57  & 0.03  & 1.22  & 0.04  & 4.43e-04& A0        & &\citet{rhee2007} \\
26990  & 85.899124  & -39.92357   & \nodata &  6.76 & 0.02  & \nodata & 6.72   & 0.01  & 6.39  & 0.04  & 0.37  & 0.05  & 2.50e-04& G0        & &\\
27259  & 86.67538   & -36.23131   & \nodata &  6.65 & 0.02  & \nodata & 6.65   & 0.01  & 5.94  & 0.04  & 0.71  & 0.04  & 1.82e-04& A3        & &\\
28186  & 89.35313   & -40.39784   & \nodata &  6.86 & 0.02  & \nodata & 6.79   & 0.01  & 6.63  & 0.05  & 0.23  & 0.05  & 1.34e-04& F5        & &\\
28230  & 89.46913   & -34.47607   & \nodata &  6.88 & 0.02  & \nodata & 6.83   & 0.01  & 5.72  & 0.03  & 1.16  & 0.04  & 7.84e-04& A8   & &\citet{rhee2007} \\
29888  & 94.4058    & -24.444433  & \nodata &  6.35 & 0.02  & \nodata & 6.23   & 0.01  & 6.14  & 0.04  & 0.21  & 0.04  & 1.29e-04& F5        & in Double system&Simbad\\
30174  & 95.234924  & -29.67068   & \nodata &  6.4  & 0.02  & \nodata & 6.08   & 0.01  & 6.03  & 0.04  & 0.37  & 0.04  & 1.73e-04& F3        & Eclipsing Binary&Simbad\\
31386  & 98.69566   & -63.262104  & \nodata &  7.05 & 0.02  & \nodata & 7.02   & 0.01  & 6.74  & 0.04  & 0.31  & 0.05  & 9.71e-05& A7       & &\\
36624  & 112.98193  & 38.896122   & \nodata &  6.42 & 0.02  & \nodata & 6.4    & 0.02  & 5.84  & 0.04  & 0.57  & 0.05  & 1.20e-04& A2        & &\\
38403  & 118.01622  & 45.933098   & \nodata &  6.02 & 0.02  & \nodata & 6.03   & 0.01  & 5.67  & 0.04  & 0.35  & 0.04  & 7.90e-05& A3         & &\\
39535  & 121.18883  & 18.842068   & \nodata &  6.28 & 0.02  & \nodata & 6.31   & 0.01  & 5.91  & 0.05  & 0.36  & 0.05  & 3.71e-05& B9        & &\\
40415  & 123.7574   & -79.318756  & \nodata &  7.02 & 0.03  & \nodata & 7.05   & 0.02  & 6.7   & 0.04  & 0.32  & 0.06  & 5.66e-05& A0        & &\\
41152  & 125.9522   & 53.21995    & \nodata &  5.25 & 0.02  & \nodata & 5.27   & 0.01  & 4.85  & 0.03  & 0.4   & 0.03  & 8.94e-05& A3        & &\citet{rhee2007} \\
42197  & 129.06482  & 42.57988    & \nodata &  6.5  & 0.02  & \nodata & 6.5    & 0.02  & 6.1   & 0.05  & 0.4   & 0.05  & 8.02e-05& A2         & &\\
42753  & 130.69264  & 31.862701   & \nodata &  6.21 & 0.02  & \nodata & 5.82   & 0.02  & 5.74  & 0.04  & 0.47  & 0.04  & 2.82e-04& F8         & Eclipsing Binary&Simbad\\
43121  & 131.73357  & 12.110076   & \nodata &  5.55 & 0.02  & \nodata & 5.56   & 0.01  & 5.06  & 0.03  & 0.49  & 0.04  & 9.12e-05& A1        & Disk &\citet{Morales2009}\\
43970  & 134.31213  & 15.322718   & \nodata &  4.87 & 0.02  & \nodata & 4.92   & 0.01  & 4.6   & 0.03  & 0.27  & 0.03  & 7.65e-05& A5      & &\citet{rhee2007} \\
46897  & 143.35873  & -22.864017  & \nodata &  5.8  & 0.02  & \nodata & 5.85   & 0.01  & 5.53  & 0.04  & 0.27  & 0.04  & 2.96e-05& B9      & &\\
46919  & 143.44405  & 62.827904   & \nodata &  5.21 & 0.02  & \nodata & 5.05   & 0.01  & 4.94  & 0.03  & 0.27  & 0.03  & 2.46e-04& G5         & Giant star&\\
47336  & 144.69057  & 10.777873   & \nodata &  6.33 & 0.03  & \nodata & 6.32   & 0.01  & 5.99  & 0.04  & 0.33  & 0.05  & 7.66e-05& A3         & in Double system&Simbad\\
47522  & 145.32094  & -23.591522  & \nodata &  4.54 & 0.02  & \nodata & 3.63   & 0.01  & 2.89  & 0.02  & 1.65  & 0.02  & 3.01e-04& B5        & Be star with excess &\citet{Touhami2011}\\
48164  & 147.26195  & 34.08553    & \nodata &  6.63 & 0.02  & \nodata & 6.66   & 0.02  & 6.19  & 0.05  & 0.44  & 0.05  & 9.67e-05& A3         & &\citet{rhee2007} \\
48541  & 148.49657  & 27.695465   & \nodata &  7.19 & 0.02  & \nodata & 7.1    & 0.02  & 6.1   & 0.04  & 1.1   & 0.05  & 3.36e-04& A0         & &\citet{rhee2007} \\
49582  & 151.83284  & -15.455298  & \nodata &  7.25 & 0.03  & \nodata & 7.14   & 0.01  & 6.23  & 0.04  & 1.02  & 0.06  & 6.29e-04& F0        & &\\
51259  & 157.0607   & -36.220123  & \nodata &  6.47 & 0.03  & \nodata & 6.47   & 0.01  & 6.24  & 0.05  & 0.24  & 0.06  & 1.27e-04& F3        & &\\
53824  & 165.1868   & 6.1015034   & \nodata &  4.61 & 0.02  & \nodata & 4.6    & 0.01  & 4.26  & 0.03  & 0.35  & 0.03  & 9.31e-05& A5      & in Double system&Simbad\\
55485  & 170.45563  & 57.074844   & \nodata &  5.99 & 0.02  & \nodata & 5.99   & 0.01  & 5.56  & 0.04  & 0.43  & 0.04  & 1.27e-04& A7       & &\citet{Morales2009}\\
55700  & 171.19781  & -22.832798  & \nodata &  6.69 & 0.02  & \nodata & 6.69   & 0.01  & 6.41  & 0.05  & 0.29  & 0.05  & 9.21e-05& A7   & &\\
59422  & 182.84108  & -3.7787178  & \nodata &  5.91 & 0.02  & \nodata & 5.93   & 0.01  & 5.52  & 0.04  & 0.39  & 0.05  & 1.95e-04& F5      &   & \\
61281  & 188.37102  & 69.78821    & \nodata &  3.82 & 0.04  & \nodata & 2.6    & 0.01  & 1.82  & 0.01  & 2.0   & 0.04  & 8.14e-04& B6     & Be star&Simbad\\
61558  & 189.19737  & -5.831851   & \nodata &  5.7  & 0.02  & \nodata & 5.71   & 0.01  & 5.18  & 0.04  & 0.52  & 0.04  & 1.17e-04& A3        & &\\
61960  & 190.47087  & 10.235843   & \nodata &  4.68 & 0.02  & \nodata & 4.7    & 0.01  & 4.27  & 0.02  & 0.41  & 0.03  & 6.91e-05& A0        & &\citet{rhee2007} \\
62576  & 192.32297  & 27.552326   & \nodata &  5.62 & 0.02  & \nodata & 5.65   & 0.02  & 5.28  & 0.03  & 0.33  & 0.04  & 6.93e-05& A2        & in Double system&Simbad\\
63942  & 196.56429  & 20.728996   & \nodata &  6.04 & 0.02  & \nodata & 5.87   & 0.01  & 5.69  & 0.04  & 0.35  & 0.04  & 1.00e-03& M0         & D or M&Simbad\\
64461  & 198.19316  & 34.528233   & \nodata &  6.77 & 0.02  & \nodata & 6.76   & 0.02  & 6.49  & 0.05  & 0.28  & 0.05  & 1.49e-04& F5         & &\\
64774  & 199.1194   & 68.40796    & \nodata &  6.27 & 0.02  & \nodata & 6.28   & 0.01  & 5.96  & 0.04  & 0.31  & 0.04  & 3.27e-05& B9      & &\\
67005  & 205.97832  & 52.064426   & \nodata &  5.99 & 0.02  & \nodata & 6.02   & 0.01  & 5.66  & 0.03  & 0.33  & 0.04  & 6.26e-05& A1      &  & \\
67495  & 207.46844  & 13.191899   & \nodata &  6.22 & 0.02  & \nodata & 6.22   & 0.01  & 5.86  & 0.04  & 0.36  & 0.04  & 7.41e-05& A2      &   & \\
67596  & 207.76878  & 34.772537   & \nodata &  6.35 & 0.02  & \nodata & 6.38   & 0.01  & 6.0   & 0.04  & 0.35  & 0.04  & 9.22e-05& A5      & & \\
67953  & 208.74294  & -8.058764   & \nodata &  5.08 & 0.03  & \nodata & 4.89   & 0.01  & 4.87  & 0.03  & 0.21  & 0.04  & 1.54e-04& F8    & in Double system&Simbad\\
69281  & 212.73259  & 15.215671   & \nodata &  6.59 & 0.03  & \nodata & 6.58   & 0.01  & 6.22  & 0.04  & 0.37  & 0.05  & 2.52e-04& G0         & C; in Double system&Simbad\\
69917  & 214.62985  & 52.03333    & \nodata &  6.56 & 0.02  & \nodata & 6.56   & 0.01  & 6.23  & 0.04  & 0.33  & 0.04  & 6.79e-05& A2         & &\\
70386  & 216.02385  & 11.246967   & \nodata &  6.07 & 0.02  & \nodata & 5.15   & 0.01  & 5.58  & 0.03  & 0.49  & 0.04  & 2.43e-04& F5         & in Double system&Simbad\\
71602  & 219.65645  & 54.853245   & \nodata &  6.41 & 0.03  & \nodata & 6.4    & 0.01  & 6.14  & 0.04  & 0.27  & 0.05  & 1.27e-04& F2         & &\\
72138  & 221.33464  & -6.7345963  & \nodata &  6.76 & 0.03  & \nodata & 6.52   & 0.01  & 6.43  & 0.05  & 0.33  & 0.06  & 2.29e-04& G0         & Eclipsing Binary&Simbad\\
72505  & 222.36974  & 19.510384   & \nodata &  6.78 & 0.02  & \nodata & 6.79   & 0.02  & 6.38  & 0.04  & 0.4   & 0.05  & 4.04e-05& B9         & C &\\
72552  & 222.49327  & 28.615833   & \nodata &  5.59 & 0.02  & \nodata & 5.57   & 0.01  & 5.05  & 0.03  & 0.53  & 0.03  & 1.33e-04& A4        & in Double system&Simbad\\
73730  & 226.07346  & 59.535053   & \nodata &  6.99 & 0.02  & \nodata & 6.96   & 0.01  & 6.63  & 0.05  & 0.36  & 0.05  & 7.36e-05& A2         & &\\
74553  & 228.497    & 43.04797    & \nodata &  6.11 & 0.02  & \nodata & 6.06   & 0.01  & 5.63  & 0.03  & 0.48  & 0.04  & 1.24e-04& A5         & &\\
75953  & 232.69203  & 34.465683   & \nodata &  6.7  & 0.02  & \nodata & 6.7    & 0.01  & 6.08  & 0.04  & 0.62  & 0.04  & 1.12e-04& A0         & &\\
76305  & 233.80867  & 65.27768    & \nodata &  7.23 & 0.02  & \nodata & 7.21   & 0.01  & 6.78  & 0.05  & 0.45  & 0.05  & 9.14e-05& A2         & C &\\
76773  & 235.12596  & 37.017002   & \nodata &  7.14 & 0.02  & \nodata & 7.13   & 0.01  & 6.41  & 0.04  & 0.74  & 0.05  & 1.47e-04& A0         & &\\
77094  & 236.12613  & 3.3701653   & \nodata &  6.31 & 0.03  & \nodata & 6.29   & 0.01  & 5.98  & 0.04  & 0.32  & 0.05  & 1.66e-04& F5         & &\\
77163  & 236.34778  & 5.4473224   & \nodata &  5.43 & 0.02  & \nodata & 5.43   & 0.01  & 4.97  & 0.03  & 0.45  & 0.03  & 8.32e-05& A1        & &\citet{rhee2007} \\
77986  & 238.87755  & 42.56615    & \nodata &  5.85 & 0.01  & \nodata & 5.65   & 0.01  & 4.99  & 0.02  & 0.86  & 0.03  & 1.16e-04& B9        & Be Star&Simbad\\
78017  & 238.95691  & 58.91171    & \nodata &  6.41 & 0.02  & \nodata & 6.39   & 0.01  & 6.07  & 0.03  & 0.34  & 0.04  & 5.86e-05& A0        & &\\
80427  & 246.2511   & 51.716564   & \nodata &  6.82 & 0.02  & \nodata & 6.83   & 0.01  & 6.55  & 0.04  & 0.27  & 0.04  & 6.63e-05& A3         & &\\
81641  & 250.1612   & 4.219815    & \nodata &  5.74 & 0.02  & \nodata & 5.78   & 0.01  & 5.33  & 0.03  & 0.4   & 0.04  & 7.43e-05& A1       & &\citet{rhee2007} \\
82587  & 253.24217  & 31.701715   & \nodata &  4.56 & 0.02  & \nodata & 4.53   & 0.01  & 4.3   & 0.03  & 0.26  & 0.03  & 1.09e-04& F0        & C; in Double system&Simbad\\
84732  & 259.8121   & 79.305435   & \nodata &  6.92 & 0.02  & \nodata & 6.91   & 0.01  & 6.65  & 0.04  & 0.26  & 0.04  & 7.62e-05& A5         & &\\
85790  & 262.95657  & 28.407436   & \nodata &  5.64 & 0.03  & \nodata & 5.63   & 0.01  & 5.25  & 0.03  & 0.39  & 0.04  & 7.16e-05& A1        & contamination &\citet{rhee2007}\\
86446  & 264.94983  & 49.779842   & \nodata &  6.58 & 0.02  & \nodata & 6.57   & 0.01  & 6.25  & 0.04  & 0.33  & 0.04  & 5.78e-05& A0         & &\\
87944  & 269.48572  & 43.417114   & \nodata &  6.86 & 0.02  & \nodata & 6.88   & 0.01  & 6.49  & 0.04  & 0.37  & 0.05  & 3.72e-05& B9         & &\\
88349  & 270.62814  & 58.62721    & \nodata &  6.58 & 0.03  & \nodata & 6.58   & 0.01  & 6.13  & 0.04  & 0.45  & 0.04  & 9.10e-05& A2         & &\\
92676  & 283.25986  & -48.360725  & \nodata &  5.85 & 0.02  & \nodata & 5.8    & 0.02  & 5.2   & 0.03  & 0.65  & 0.04  & 1.43e-04& A2        & &\\
93542  & 285.77847  & -42.094994  & \nodata &  4.75 & 0.02  & \nodata & 4.72   & 0.01  & 3.79  & 0.03  & 0.96  & 0.03  & 2.44e-04& A0       & &\citet{rhee2007} \\
94140  & 287.44083  & 65.97845    & \nodata &  6.26 & 0.02  & \nodata & 6.19   & 0.01  & 5.45  & 0.03  & 0.81  & 0.03  & 1.72e-04& A0       & &\\
95261  & 290.71326  & -54.42373   & \nodata &  5.01 & 0.03  & \nodata & 4.73   & 0.01  & 3.25  & 0.02  & 1.76  & 0.04  & 1.53e-03& A0       & &\citet{rhee2007} \\
95270  & 290.74548  & -54.53785   & \nodata &  5.91 & 0.03  & \nodata & 5.89   & 0.01  & 3.96  & 0.02  & 1.95  & 0.04  & 7.10e-03& F5/F6     & &\citet{rhee2007} \\
99892  & 304.01157  & -16.295567  & \nodata &  6.72 & 0.02  & \nodata & 6.68   & 0.02  & 5.93  & 0.05  & 0.79  & 0.05  & 1.67e-04& A0      & &\\
100787 & 306.5198   & -46.659958  & \nodata &  6.15 & 0.02  & \nodata & 6.12   & 0.01  & 5.77  & 0.04  & 0.38  & 0.05  & 1.38e-04& A9   & &\\
102419 & 311.31223  & -15.79827   & \nodata &  6.12 & 0.03  & \nodata & 6.14   & 0.02  & 5.8   & 0.05  & 0.33  & 0.05  & 1.43e-04& F2/F3  & &\\
103048 & 313.17328  & -53.273277  & \nodata &  6.56 & 0.02  & \nodata & 6.54   & 0.01  & 6.05  & 0.05  & 0.51  & 0.05  & 2.56e-04& F5/F6     & &\\
105169 & 319.5675   & -75.346664  & \nodata &  6.54 & 0.02  & \nodata & 6.55   & 0.01  & 6.05  & 0.04  & 0.49  & 0.05  & 8.97e-05& A1        & &\\
106741 & 324.33783  & -18.440924  & \nodata &  6.18 & 0.02  & \nodata & 6.18   & 0.01  & 5.93  & 0.05  & 0.25  & 0.05  & 1.31e-04& F3/F5    & &\citet{rhee2007} \\
106783 & 324.4317   & 6.6183963   & \nodata &  6.11 & 0.02  & \nodata & 6.13   & 0.01  & 5.77  & 0.04  & 0.34  & 0.04  & 7.08e-05& A2        & in Double system&Simbad\\
107336 & 326.0995   & -4.730885   & \nodata &  6.41 & 0.02  & \nodata & 6.39   & 0.02  & 6.13  & 0.05  & 0.28  & 0.05  & 6.15e-05& A2         & &\\
107585 & 326.8533   & -4.6086183  & \nodata &  6.68 & 0.02  & \nodata & 6.64   & 0.02  & 5.77  & 0.05  & 0.91  & 0.05  & 2.62e-04& A2         & &\\
107596 & 326.90918  & -5.91684    & \nodata &  5.63 & 0.03  & \nodata & 5.62   & 0.01  & 5.22  & 0.04  & 0.41  & 0.05  & 1.22e-04& A7        & &\\
107919 & 327.9661   & 11.091197   & \nodata &  6.05 & 0.03  & \nodata & 6.07   & 0.01  & 5.55  & 0.04  & 0.5   & 0.05  & 1.31e-04& A5         & &\\
109230 & 331.9346   & -28.147606  & \nodata &  6.47 & 0.03  & \nodata & 6.49   & 0.02  & 6.2   & 0.05  & 0.27  & 0.05  & 1.12e-04& Fm         & &\\
110739 & 336.52835  & -5.1778803  & \nodata &  6.56 & 0.02  & \nodata & 6.55   & 0.02  & 6.14  & 0.05  & 0.41  & 0.05  & 1.55e-04& F0         & &\\
110786 & 336.67352  & -11.22817   & \nodata &  6.32 & 0.02  & \nodata & 6.34   & 0.01  & 5.83  & 0.05  & 0.49  & 0.05  & 1.10e-04& A3         & &\\
112542 & 341.92813  & -14.056405  & \nodata &  5.73 & 0.02  & \nodata & 5.74   & 0.01  & 5.15  & 0.03  & 0.58  & 0.04  & 6.09e-05& B9        & in Double system&Simbad\\
114822 & 348.89276  & -3.4963725  & \nodata &  5.4  & 0.02  & \nodata & 5.43   & 0.01  & 5.1   & 0.03  & 0.31  & 0.04  & 7.17e-05& A3        & &\\
116431 & 353.90048  & 8.382718    & \nodata &  6.4  & 0.02  & \nodata & 6.41   & 0.01  & 5.01  & 0.03  & 1.4   & 0.04  & 1.49e-03& F0         & &\citet{rhee2007} \\
117481 & 357.33157  & -27.854132  & \nodata &  5.7  & 0.03  & \nodata & 5.71   & 0.01  & 5.44  & 0.04  & 0.26  & 0.05  & 1.54e-04& F6/F7     & &\\
118133 & 359.4425   & 11.4744005  & \nodata &  6.59 & 0.02  & \nodata & 6.58   & 0.02  & 6.11  & 0.05  & 0.48  & 0.05  & 4.87e-05& B9        & in Double system&Simbad\\
118322 & 359.9788   & -65.57708   & \nodata &  4.6  & 0.02  & \nodata & 4.57   & 0.01  & 4.31  & 0.02  & 0.29  & 0.03  & 3.13e-05& B9       & Be Star&Simbad\\
\enddata
\tablecomments{C: Contaminated stars from W4 band images;\\
    D or M: Double or multiple star;\\
    Disk: IR excess stars with disk;\\
    Double?: May be the double system;\\
    The editor can get the full version of this Table from author.}
\end{deluxetable}

\clearpage
\begin{deluxetable}{rrr}
\tablecolumns{3}
\tablewidth{0pc}
\tablecaption{22 $\mu$m Excess Detect rate of Main-Sequence Stars 
\label{tab_detec}}
\tablehead{
    \colhead{SpType} & \colhead{Detection} & \colhead{Total} \\
}
\startdata
B & 20 & 120  \\
A & 76 & 617  \\
F & 36 & 1220 \\
G & 4  & 531  \\
K & 1  & 142  \\
M & 3  & 19   \\
\enddata
\tablecomments{The total number only contains those with W4 $S/N\ge20$ and low 
IRAS 100 $\mu$m background level.}
\end{deluxetable}

\appendix
\section{Gallery of images and SEDs (The first 3 stars)
\label{appendixc}}
\begin{figure}[h!tb]
\begin{center}
\includegraphics[width=0.33\textwidth]{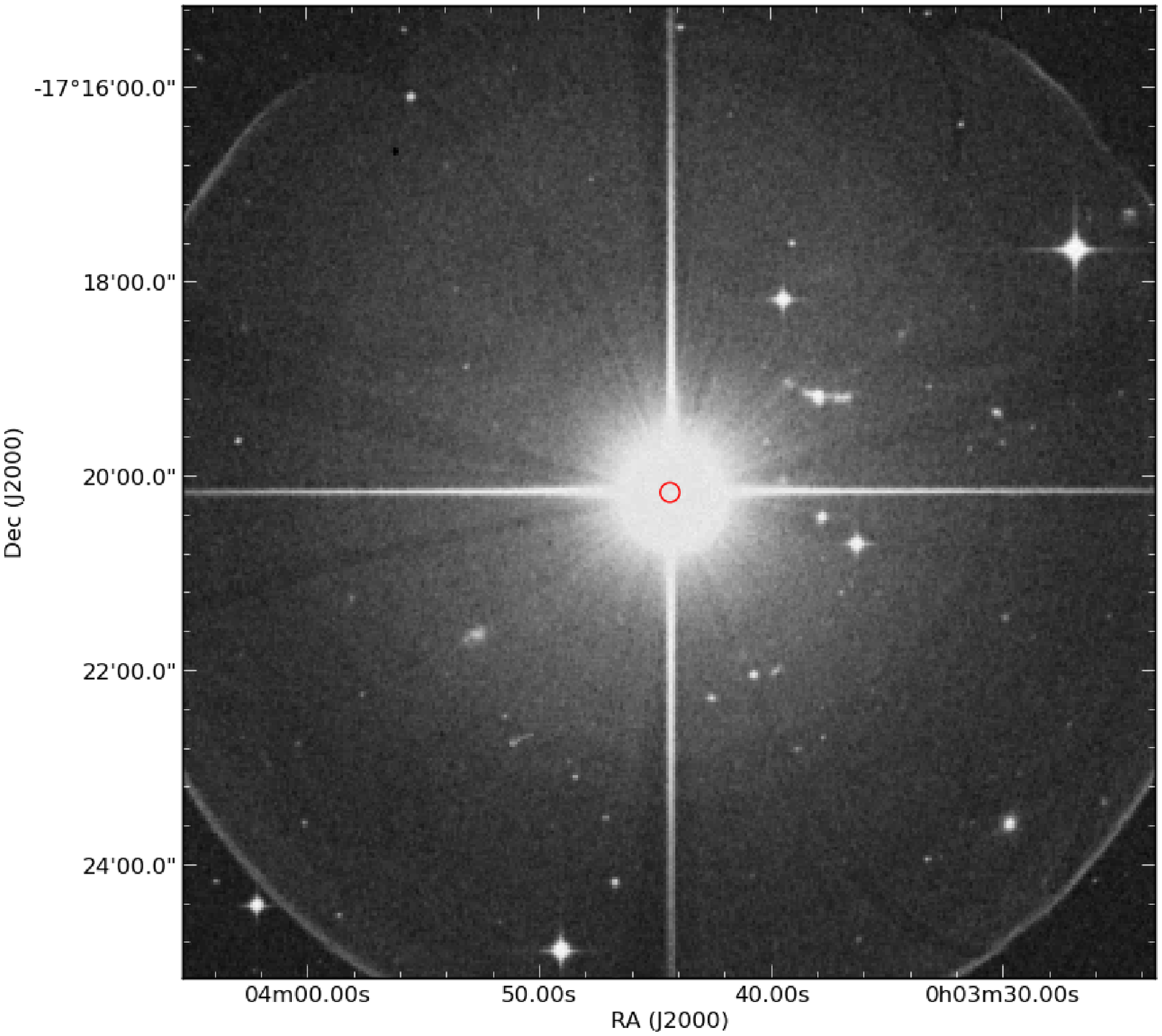}%
\includegraphics[width=0.28\textwidth]{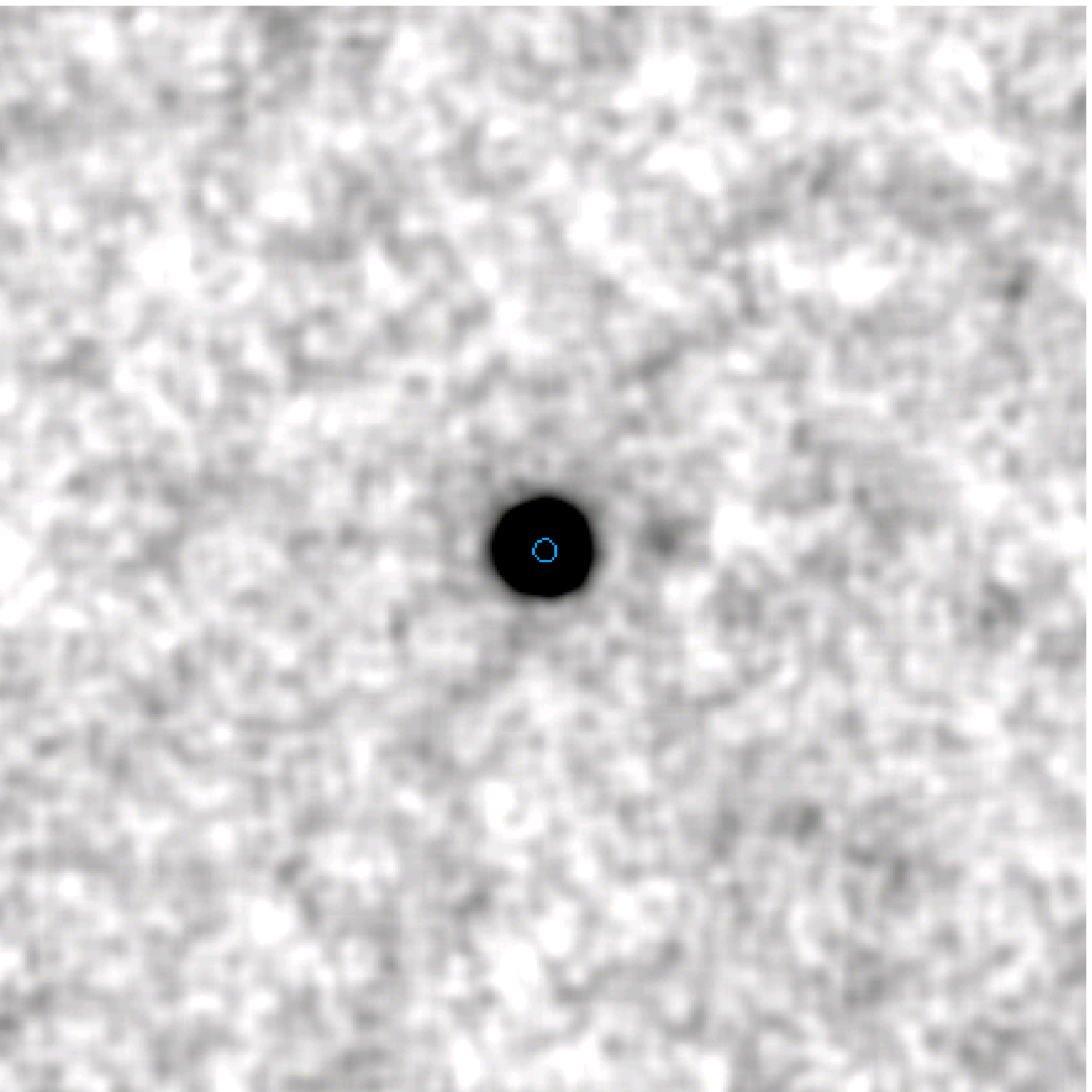}%
\includegraphics[width=0.33\textwidth]{hip301}%
\end{center}
\end{figure}
\begin{figure}[h!tb]
\begin{center}
\includegraphics[width=0.33\textwidth]{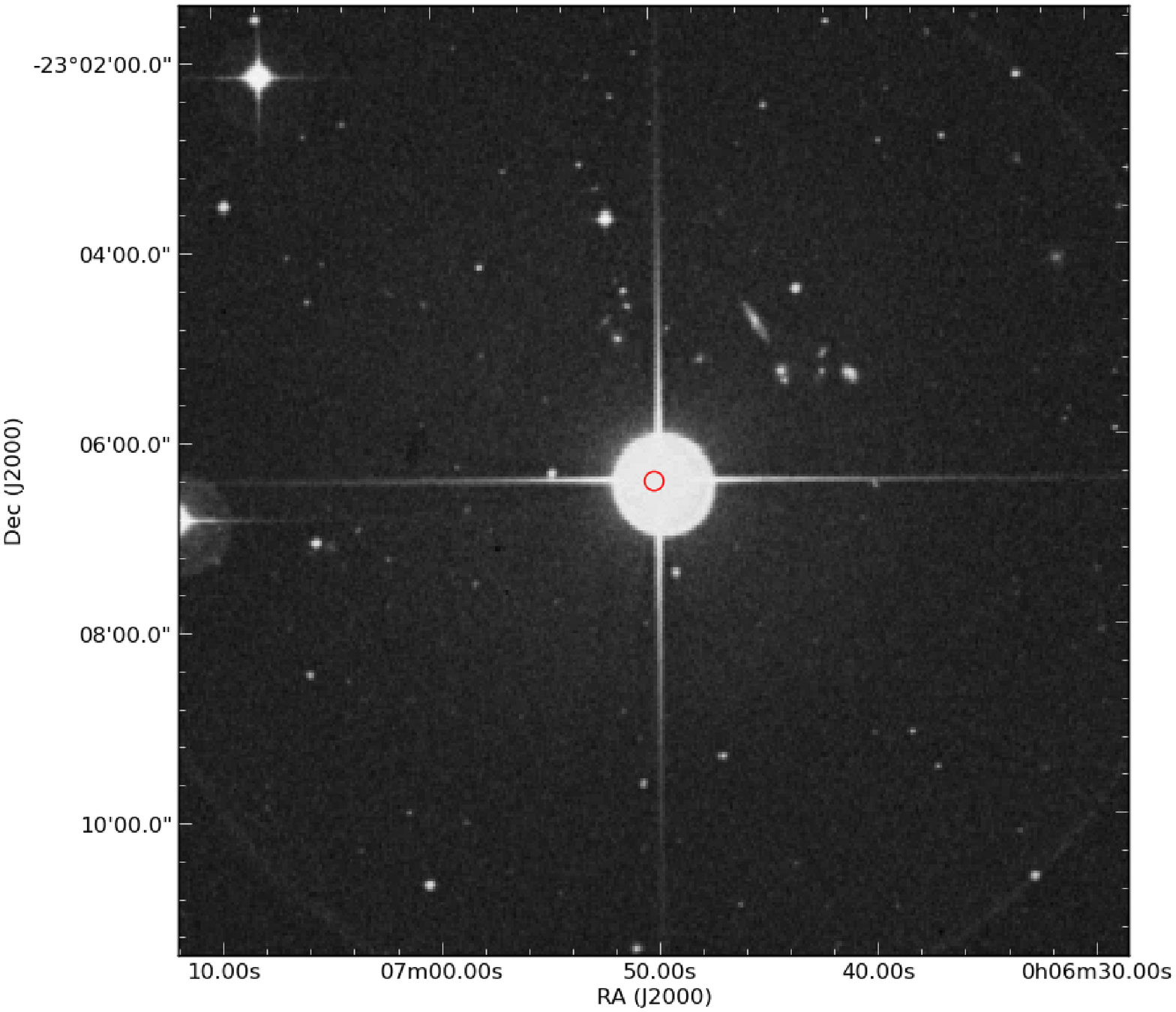}%
\includegraphics[width=0.28\textwidth]{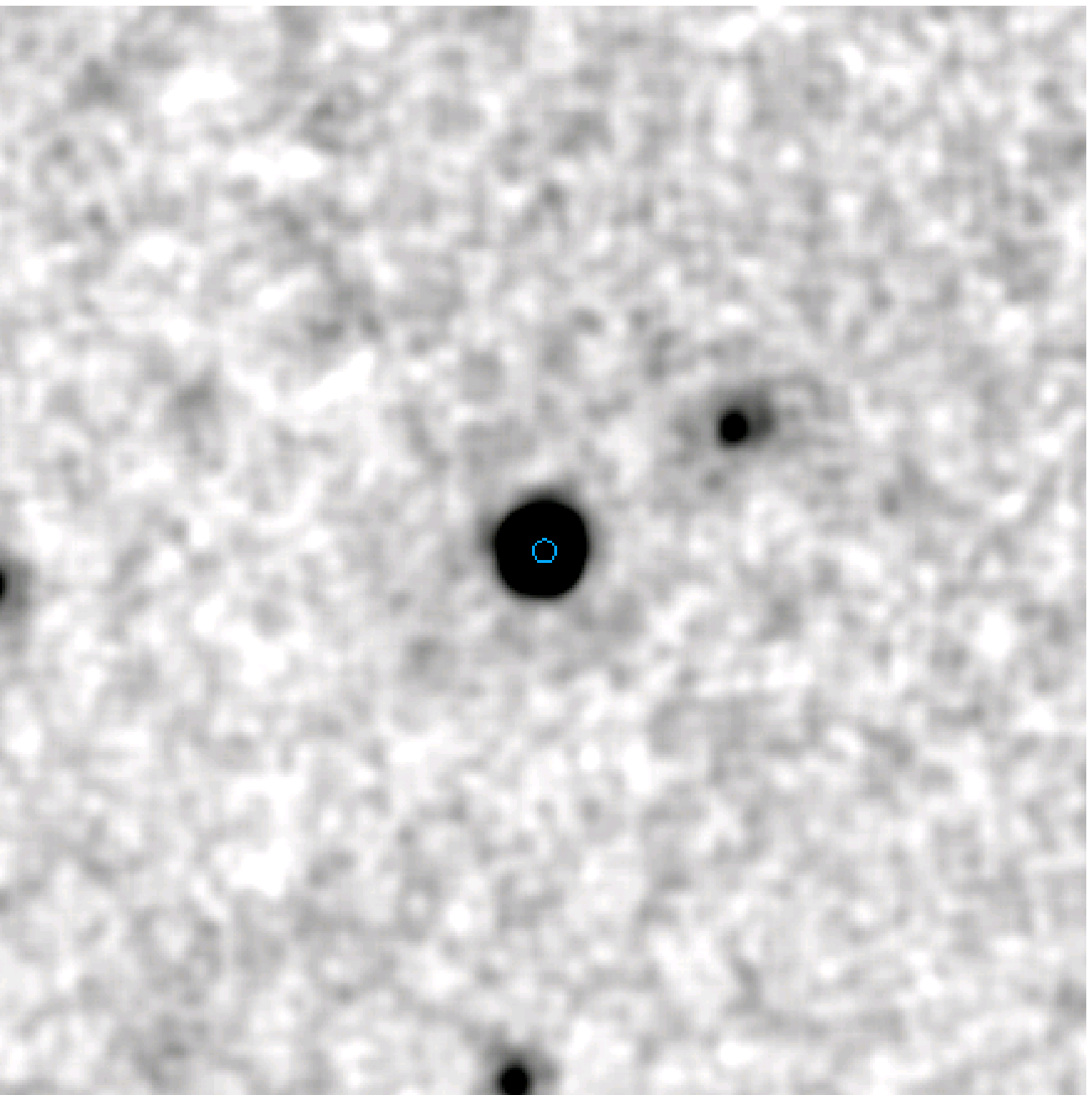}%
\includegraphics[width=0.33\textwidth]{hip560}%
\end{center}
\end{figure}
\begin{figure}[h!tb]
\begin{center}
\includegraphics[width=0.33\textwidth]{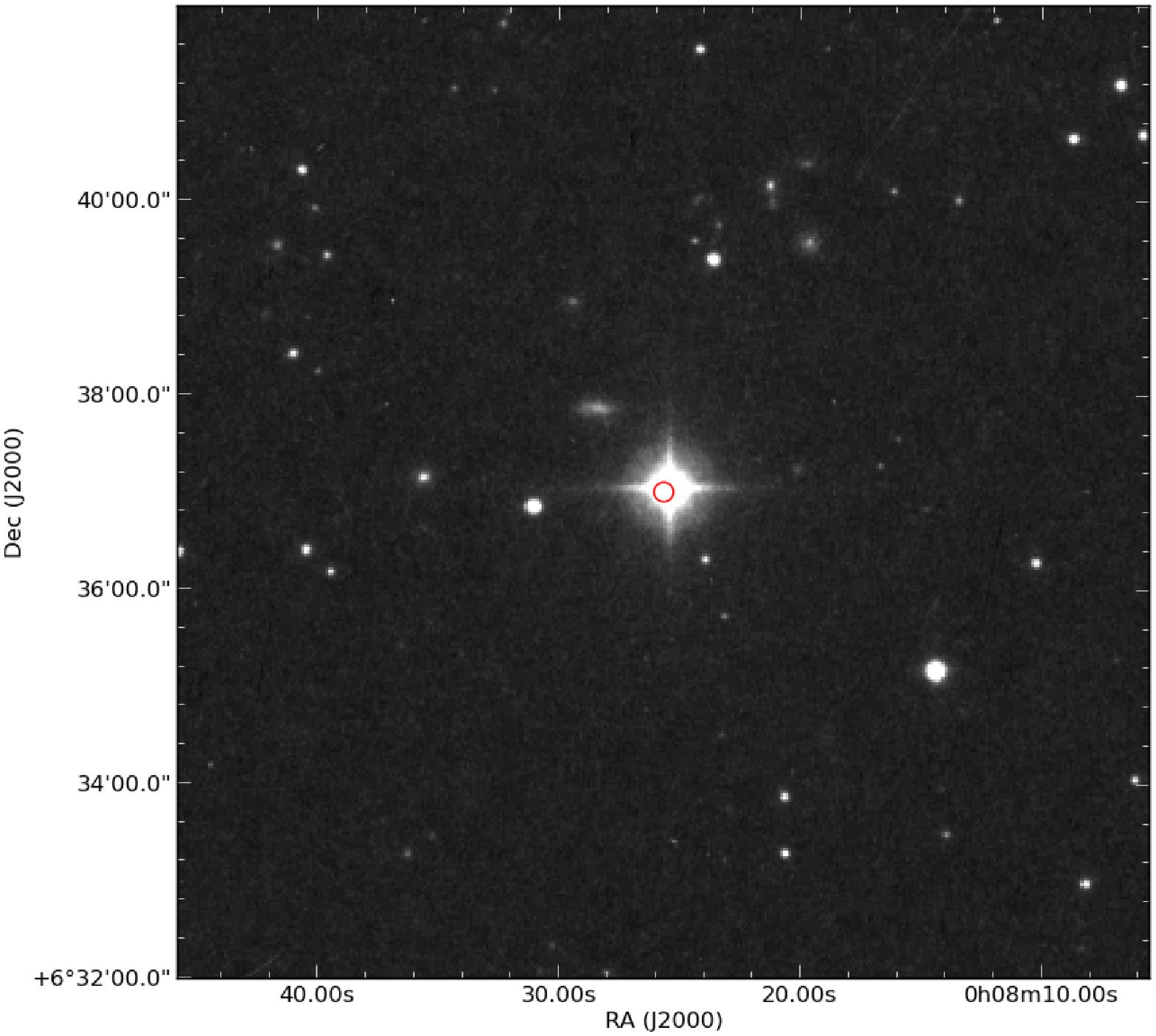}%
\includegraphics[width=0.28\textwidth]{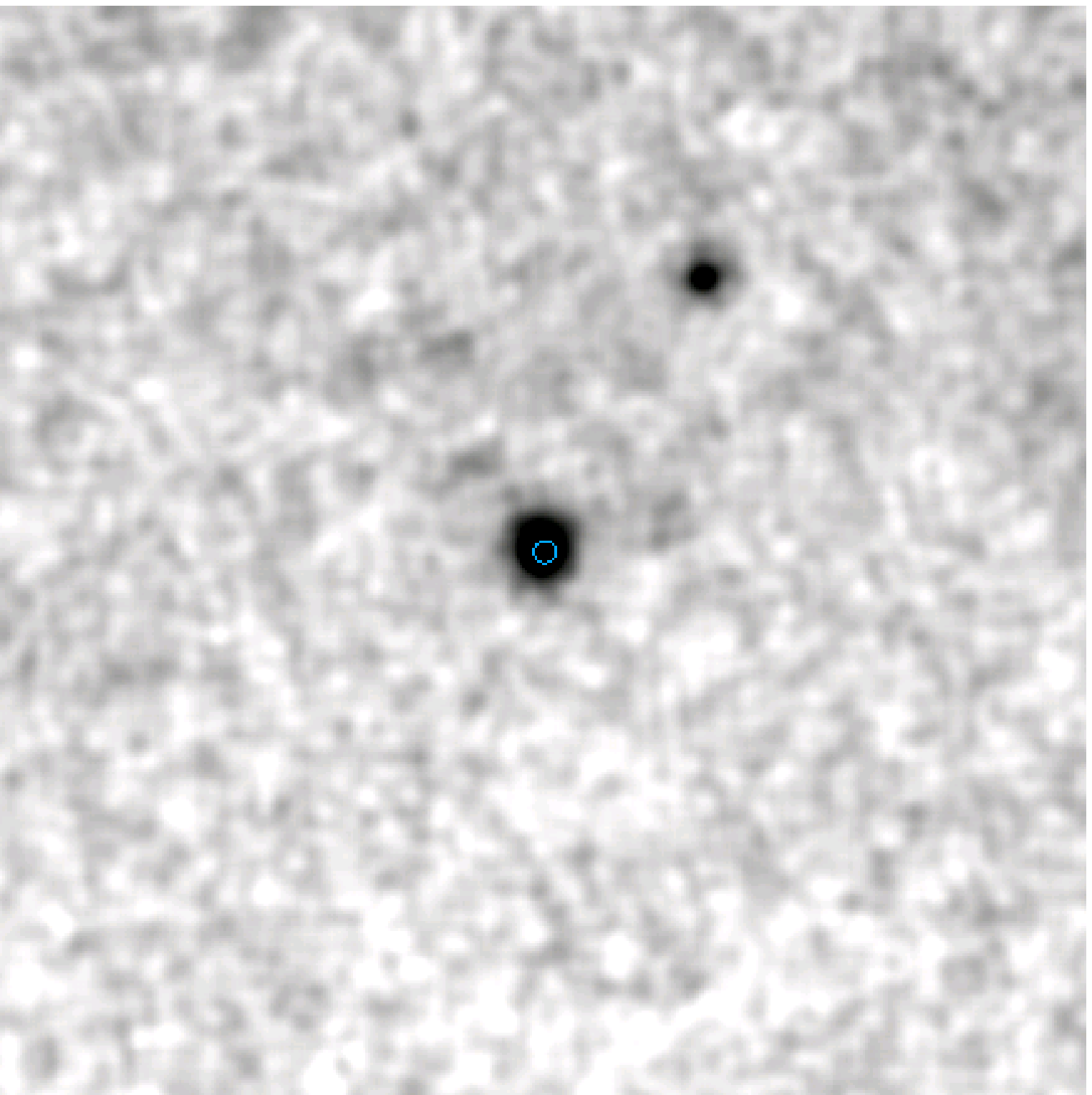}%
\includegraphics[width=0.33\textwidth]{hip682}%
\end{center}
\caption{The optical images and SEDs of the first 3 stars. From top to bottom, 
    the name of stars are hip301, hip560, hip682, respectively.
    The editor can get the full version of this gallery from author.}
\end{figure}

\end{document}